\documentclass[prb,aps,amssymb,twocolumn,showpacs,floatfix]{revtex4}
\usepackage[dvips]{graphicx}
\def\be{\begin{equation}}
\def\ee{\end{equation}}
\def\bea{\begin{eqnarray}}
\def\eea{\end{eqnarray}}

\begin{document}

\title{Transport of interacting electrons through a double barrier in
quantum wires}

\author{D. G. Polyakov$^{1,*}$}
\author{I. V. Gornyi$^{1,2,*}$}
\affiliation{$^{1}$Institut f\"ur Nanotechnologie,
Forschungszentrum Karlsruhe, 76021 Karlsruhe, Germany \\
$^{2}$Institut f\"ur Theorie der kondensierten Materie, Universit\"at
Karlsruhe, 76128 Karlsruhe, Germany}


\begin{abstract} We generalize the fermionic renormalization group
method to describe analytically transport through a double barrier
structure in a one-dimensional system.  Focusing on the case of weakly
interacting electrons, we investigate thoroughly the dependence of the
conductance on the strength and the shape of the double barrier for
arbitrary temperature $T$. Our approach allows us to systematically
analyze the contributions to renormalized scattering amplitudes from
different characteristic scales absent in the case of a single
impurity, without restricting the consideration to the model of a
single resonant level.  Both a sequential resonant tunneling for high
$T$ and a resonant transmission for $T$ smaller than the resonance
width are studied within the unified treatment of transport through
strong barriers.  For weak barriers, we show that two different
regimes are possible.  Moderately weak impurities may get strong due
to a renormalization by interacting electrons, so that transport is
described in terms of theory for initially strong barriers. The
renormalization of very weak impurities does not yield any peak in the
transmission probability; however, remarkably, the interaction gives
rise to a sharp peak in the conductance, provided asymmetry is not too
high. \end{abstract}


\maketitle

\section{Introduction}
\label{I}

Effects related to the Coulomb interaction between electrons become
increasingly prominent in systems of lower spatial dimensionality as
their size is made smaller. Recent experimental progress in controlled
preparation of nanoscale devices has led to a revival of interest in
the transport properties of one-dimensional (1D) quantum wires. Owing
to the particular geometry of the Fermi surface, systems of
dimensionality one are unique in that the Coulomb correlations in 1D
change a noninteracting picture completely and thus play a pivotal
role in low-temperature physics. 

A remarkable example of a correlated 1D electron phase is the
Luttinger-liquid model (for a review see, e.g.,
Refs.~\onlinecite{solyom79,voit94}). In this model, arbitrarily weak
interactions ruin the conventional Fermi liquid phenomenology by
essentially modifying low-energy excitations across the Fermi
surface. As a result, the tunneling density of states develops
power-law singularities on the Fermi surface. Moreover, interactions
between oppositely moving electrons generate charge- and spin-density
wave correlations that lead to striking transport properties of a
Luttinger liquid in the presence of impurities. In particular, even a
single impurity yields a complete pinning of a Luttinger liquid with
repulsive interactions, which results in a vanishing conductance at
zero temperature.\cite{kane92,furusaki93a} In addition to quantum
wires, largely similar ideas apply to edge modes in a Hall bar
geometry in the fractional quantum Hall regime, which are thought to
behave as spatially separated chiral Luttinger liquids.\cite{wen91}

Evidence has recently emerged pointing towards the existence of the
Luttinger liquid in metallic single-wall carbon
nanotubes.\cite{bockrath99,yao99} The Luttinger liquid behavior was
observed via the power-law temperature and bias-voltage dependence of
the current through tunneling contacts attached to the
nanotubes. Further technological advances have made possible the
fabrication of low-resistance contacts between nanotubes and metallic
leads (see, e.g.,
Refs.~\onlinecite{yao00,nygard00,bockrath01,krupke02} and references
therein). These recent developments have paved the way for systematic
transport measurements in Luttinger liquids with impurities.

In this paper, we study electron transport through a double barrier in
a 1D liquid. In the 1D geometry, two impurities in effect create a
quantum dot inside the system.  Resonant tunneling through the two
impurities is a particularly attractive setup to investigate the
correlated transport in an inhomogeneous Luttinger liquid. Due to the
resonant behavior of the current, the interplay of Luttinger-liquid
correlations and impurity-induced backscattering should be more easily
accessible to transport measurements. Two striking experimental
observations have been reported recently. In
Ref.~\onlinecite{auslaender00}, a resonant structure of the
conductance of a semiconductor single-mode quantum wire was
attributed to the formation (with reduction of electron density by
changing gate voltage) of a single disorder-induced quantum dot. In
Ref.~\onlinecite{postma01}, two barriers were created inside a carbon
nanotube in a controlled way with an atomic force microscope. In both
cases, the amplitude of a conductance peak $G_p$ as a function of
temperature $T$ showed power-law behavior $G_p\propto {T}^{-\gamma}$
with the exponent $\gamma$ noticeably different from $\gamma=1$. The
latter is the value of $\gamma$ expected in the absence of
interactions (for a review, see
Refs.~\onlinecite{restun2,restun3}) provided $T$ lies in the
range $\Gamma\ll {T}\ll \Delta$, where $\Gamma$ is the width of a
resonance in the transmission coefficient and $\Delta$ is the
single-particle level spacing. The width of a conductance peak $w$
followed a linear temperature dependence $w\propto {T}$ in both
experiments.

On the theoretical side, resonant tunneling in a Luttinger liquid was studied
previously in a number of
papers.\cite{kane92,furusaki93,sassetti95,maurey95,furusaki98,braggio00} In
particular, the width $\Gamma\propto {T}^{\alpha_e}$ was
shown\cite{furusaki93,furusaki98} to shrink with decreasing temperature.  The
exponent $\alpha_e$ depends on the strength of interaction and describes
tunneling into the end of a semi-infinite liquid. The dimensionless peak
conductance (in units of $e^2/h$) obeys $G_p\sim \Gamma/{T}$ in the above
range of $T$, which indeed leads to a smaller value of $\gamma=1-\alpha_e$.
The reduced exponent $\gamma$ reported in Ref.~\onlinecite{auslaender00} was
positive (and different for different conductance peaks, in the range
$\gamma\sim 0.5-0.8$), whereas in Ref.~\onlinecite{postma01} the reported
value of $\gamma\simeq -0.7$ was negative. More specifically, in
Ref.~\onlinecite{postma01}, the conductance as a function of the gate voltage
showed certain traces of periodicity characteristic to the Coulomb blockade
regime.\cite{restun2,restun3,cb} Surprisingly, both the amplitude $G_p$ and
the width $w$ were reported to vanish with decreasing $T$, in sharp contrast
to the noninteracting case.  While such behavior is known to be possible for
very strong repulsive interaction,\cite{furusaki93,furusaki98} the required
strength of interaction would then be much larger than expected and indeed
reported (see Refs.~\onlinecite{bockrath99,yao99,eggergo98,kane97} and
references therein) in carbon nanotubes. Roughly a doubling (or even a larger
factor) of the expected\cite{eggergo98,kane97,bockrath99,yao99} exponent
$\alpha_e$, which is $\alpha_e\sim 0.6-1.0$, would be necessary to fit the
experimental data. Although the modeling\cite{eggergo98,kane97} of carbon
nanotubes as a four-channel Luttinger liquid has its share of complications,
the observations\cite{postma01} appear to present a puzzle.

It was suggested in Ref.~\onlinecite{postma01} that a certain novel mechanism
of ``correlated tunneling" dominates over the conventional sequential
tunneling for ${T}\gg \Gamma$, leading to a doubling of the exponent $\alpha$,
which might explain the experiment. In the subsequent
works\cite{thorwart02,thorwart021} the basic ideas relevant to the resonant
tunneling have been questioned. In particular, the lowest-order contribution
to the resonance peak conductance $G_p$ for $\Delta\gg T\gg \Gamma$ has been
argued\cite{thorwart02,thorwart021} to come from processes of second order in
the end-tunneling density of states. While being a characteristic feature of
the cotunneling regime far in the wings of the conductance peak, this,
however, disagrees with the sequential tunneling picture inside the
conductance peak.\cite{furusaki98,furusaki93} This also poses a problem with
the persistence\cite{kane92} of perfect transmission through symmetric
barriers in the case of weak interaction. Moreover, the main
suggestion\cite{thorwart02} is that taking a finite-range (but falling off
fast beyond this range) interaction changes the conventional picture
completely, as compared to the standard treatment of a zero-range interaction
in the Luttinger liquid model. According to Ref.~\onlinecite{thorwart02},
higher-order tunneling processes in combination with the effects of a non-zero
range of interaction dominate the dependence of $G_p$ on $T$ even for ${T}\gg
\Gamma$.  A nearly perfect agreement with the experiment data\cite{postma01}
has been claimed (for a range of interaction far smaller than the distance
between the barriers). However, no explicit formula for the conductance has
been given in Ref.~\onlinecite{thorwart02}, while we see no ground for the
low-energy long-distance physics to be essentially different if the radius of
interaction is made finite.

In another recent attempt\cite{kleimann02} to explain the
experiment,\cite{postma01} the observed power law was attributed to the
contact resistance. Namely it was noted that the resistance of tunneling
contacts to the leads, $R_c$, and the resistance of the quantum dot add up (if
one applies Kirchhoff's law), so that the anomalous $T$ dependence in
Ref.~\onlinecite{postma01} might be explained if $R_c\gg G_p^{-1}$, where
$G_p$ is understood as the resonance peak conductance of the dot. This is a
legitimate proposition, although the measured contact resistance in
Ref.~\onlinecite{postma01} was reported to be relatively
low.\cite{postma01,thorwart02}

It is thus desirable to examine the resonant tunneling in a Luttinger
liquid in a broad range of temperature down to ${T}=0$ and for various
parameters of the barriers. Our purpose in this paper is to analyze
transport through a double barrier of arbitrary strength, strong or
weak, symmetric or asymmetric, within a general framework of an
analytical method applicable to all these situations. There are a
variety of techniques to construct the low-energy transport
theory.\cite{voit94,gogolin98,fisher96} The method we develop here is
valid for weak interaction and is based on the renormalization group
(RG) approach of Refs.~\onlinecite{matveev93,yue94}, which was applied
earlier in a variety of contexts.\cite{kane94,tsai02,nazarov02} One of
the appeals of this kind of theory is that it allows one to treat weak
and strong scatterers on an equal footing, which is technically
significantly less straightforward in the bosonization
method.\cite{gogolin98} Thus it would be of interest to apply this
method to a 1D mesoscopic interacting system with many impurities
(previously, the problem of transport in a disordered Luttinger liquid
was studied by perturbative in disorder methods based on bosonic field
theories in, e.g.,
Refs.~\onlinecite{giamarchi88,furusaki93a,maslov95}). A double barrier
is the simplest ``many-impurity" system which exhibits effects
essential to the physics of disordered interacting 1D liquids.

Very recently, a generalization of the RG approach (initially developed for a
structureless barrier\cite{matveev93,yue94}) has been proposed in
Ref.~\onlinecite{nazarov02} for the model of a single impurity with an
energy-dependent scattering matrix. This model is relevant to the resonant
tunneling of weakly interacting electrons through a double barrier and
accounts properly for the interaction processes within an energy band of width
$\Delta$ around the Fermi level.  In particular, a non-monotonic behavior of
the conductance peak as a function of $T$, caused by left/right asymmetry of
scattering amplitudes, was investigated in Ref.~\onlinecite{nazarov02} within
the single-resonance model. The fermionic RG\cite{nazarov02} has been shown to
be in accord with the results\cite{kane92,furusaki93,furusaki98} for the
resonant tunneling in Luttinger liquids, obtained earlier by different
methods.

However, to describe microscopically the spatial structure of a system of two
or more impurities, a more systematic analysis is needed. Specifically, one
has to develop an approach that would include contributions to the
renormalized scattering amplitudes due to interaction processes involving
energy transfers larger than $\Delta$. Also, the single-resonance
model\cite{nazarov02} does not describe contributions from tunneling through
multiple levels inside the quantum dot (multi-level quantum dots represent a
typical experimental situation). Finally, it would be interesting to study
transport through weak impurities (and through strongly asymmetric structures)
for which the bare transmission coefficient exhibits no pronounced resonant
structure.  All this adds to our motivation to study the resonant tunneling
through a double barrier by generalizing the RG approach of
Refs.~\onlinecite{matveev93,yue94}.

The RG approach enables us to investigate in detail the resonant transport of
weakly interacting spinless electrons. Within the fermionic RG approach, we
confirm earlier results\cite{kane92,furusaki93,furusaki98} obtained within
bosonic field theories. We examine the conductance through a double barrier
for arbitrary strength and an arbitrary shape of the barrier, not restricting
ourselves to the model of a single resonant level. In particular, we
demonstrate the existence of narrow conductance peaks for two weak impurities,
which is in sharp contrast to the noninteracting case. We do not find any
trace of the correlated tunneling mechanism proposed in
Refs.~\onlinecite{postma01,thorwart02,thorwart021}. We also clarify the
relationship between the RG method\cite{matveev93,yue94} and Hartree-Fock (HF)
approaches.

The paper is organized as follows. First, in Sec.~\ref{IIa}, we
briefly outline the fermionic RG approach to transport through a
single impurity. In Sec.~\ref{IIb}, we discuss a HF treatment of
transmission through a single impurity and show its inadequacy to the
problem. In Sec.~\ref{III}, we turn to a double barrier. We start with
a perturbative expansion in Sec.~\ref{IIIa} and derive the RG equation
for transport through the double barrier in Sec.~\ref{IIIb}. We then
analyze contributions to scattering amplitudes from different energy
scales, compared to the level spacing inside the quantum dot and the
resonance width, in Secs.~\ref{IIIc}--\ref{IIIe}. In Sec.~\ref{IIIf},
we concentrate on the case of weak impurities. Finally, in
Sec.~\ref{IV} we calculate the amplitude and the shape of the
resonance conductance peaks. 

\section{Single impurity}
\label{II}

\subsection{Renormalization group: Basic results}
\label{IIa}

We begin with a brief description of transport through a single
structureless impurity in the spirit of the RG
approach\cite{matveev93,yue94}. Without interaction, the impurity is
characterized by a transmission coefficient $t_0$ and reflection
coefficients $r_{L0}$ and $r_{R0}$, from the left and from the right
respectively (we put the impurity at the center of coordinates,
$x=0$). Suppose that the energy dependence of the bare scattering
matrix can be neglected far from the boundaries of an energy band
$(-D_0,D_0)$ around the Fermi level. The energy scale $D_0$ serves as
the ultraviolet cutoff of RG transformations and, physically, is of
the order of $v_F/d$ (throughout the paper we put $\hbar=1$) or the
Fermi energy $\epsilon_F$, whichever is smaller.\cite{cutoff} Here $d$
is the radius of interaction and $v_F$ is the Fermi velocity. Deep
inside the band $(-D_0,D_0)$, we linearize the energy spectrum around
the Fermi level. The differential RG equations\cite{matveev93,yue94}
read
\be
\partial t/\partial {\cal L}=-\alpha t\,{\rm R}~,\quad 
\partial r_{L,R}/\partial
{\cal L}=\alpha r_{L,R}{\rm T}~,
\label{1}
\ee
where ${\cal L}=\ln (D_0/|\epsilon|)$, the energy $\epsilon$ is
measured from the Fermi level, the transmission probability ${\rm T}=|t|^2$,
and ${\rm R}=1-{\rm T}$. The boundary conditions at ${\cal L}=0$ set the
scattering amplitudes at their noninteracting values $t_0$, $r_{L0}$,
and $r_{R0}$. Throughout the paper we consider spinless electrons, for
which the interaction constant is 
\be
\alpha=(V_f-V_b)/2\pi v_F~,
\label{2}
\ee
where $V_f$ and $V_b$ are the Fourier transforms of a pairwise interaction
potential yielding forward ($V_f$) and backward ($V_b$) scattering. The
forward scattering does not lead to transitions between two branches of right-
and left-movers, whereas the backscattering does. We assume that $\alpha>0$.

Note that, for spinless electrons, the interaction-induced backward scattering
and forward scattering relate to each other as direct and exchange processes,
so that the backscattering only appears in the combination $V_f-V_b$ and thus
merely redefines parameters of the Luttinger model (formulated
\cite{solyom79,voit94} in terms of forward-scattering amplitudes only). In
particular, the backscattering does not lead to any RG flow for $\alpha$. For
spinful electrons this is valid only to one-loop order.\cite{solyom79,voit94}

It is also worth mentioning that for a point interaction $V_f=V_b$, so that
$\alpha=0$, hence for spinless electrons one has to start with a finite-range
interaction. However, the RG flow for the scattering matrix (\ref{1}) occurs
for $|\epsilon|\alt v_F/d$ and is governed solely by the constant $\alpha$. It
follows that on low-energy scales one can effectively consider the interaction
as local, $V_{\rm eff}(x-x')=2\pi \alpha v_F\delta(x-x')$, and formally deal
exclusively with forward scattering. A non-zero range of interaction for
$k_Fd\gg 1$ can manifest itself only in the boundary conditions to
Eqs.~(\ref{1}) at $|\epsilon|\sim D_0=v_F/d$ and therefore does not affect the
singular behavior of the renormalized scattering matrix at $\epsilon\to 0$. We
assume that the Coulomb interaction between electrons is screened by external
charges (e.g., by metallic gates, in which case $d$ is given by the distance
to the gates) and that a resulting $\alpha\ll 1$. For a treatment of the
unscreened Coulomb interaction, see, e.g.,
Refs.~\onlinecite{schulz93,maurey95}.

\begin{figure}[ht]
\centerline{
\includegraphics[width=8cm]{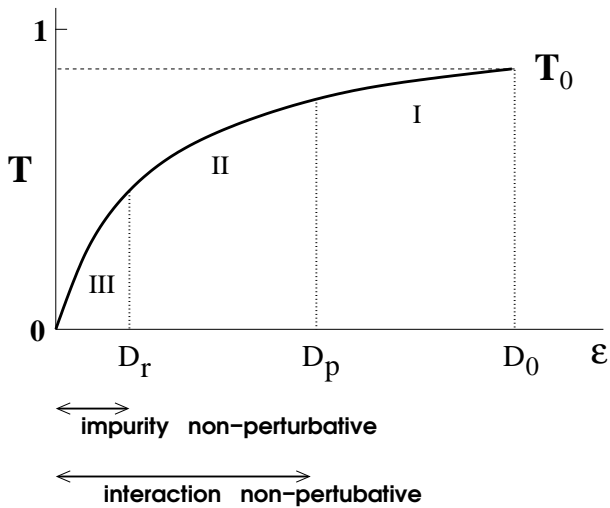}} 
\caption{\label{sca} 
  Schematic behavior of the transmission coefficient ${\rm T}(\epsilon)$ for
  weak interaction and a single weak impurity with a bare coefficient ${\rm
    T}_0\simeq 1$ in three different regions, according to
  Eqs.~(\ref{3}),(\ref{4}): (I) a small logarithmic correction $\delta {\rm
    T}$; (II) power-law scaling of the correction $\delta {\rm T}$; (III)
  power-law vanishing of ${\rm T}$. The Fermi level is at $\epsilon=0$. }
\end{figure}

Integration of Eqs.~(\ref{1}) gives\cite{matveev93,yue94}
\be
{{\rm R}\over {\rm T}}={{\rm R}_0\over 
{\rm T}_0}\left(D_0\over |\epsilon|\right)^{2\alpha}~.
\label{3}
\ee The phases of the scattering amplitudes are not affected by the
renormalization. Equations (\ref{1}) are equivalent to a one-loop
renormalization, so that Eq.~(\ref{3}) is valid to first order in interaction
$\sim {\it O}(\alpha)$ in the exponent of the power-law scaling. As follows
from Eq.~(\ref{3}), whatever the initial values of ${\rm T}_0$, at $\alpha >0$
they all flow to the fixed point of Eqs.~(\ref{1}) at zero
transmission,\cite{kane92} ${\rm T}=0$ at $\epsilon=0$, see Fig.~\ref{sca}. In
the limits of a weak impurity (both ${\rm R}_0\ll 1$ and ${\rm R}\ll 1$) and a
strong tunneling barrier (${\rm T}_0\ll 1$), Eq.~(\ref{3}) coincides with the
RG results obtained by bosonization,\cite{kane92} provided $\alpha\ll 1$.
Equation (\ref{3}) gives the transmission probability for electrons with
energy $\epsilon$ at temperature ${T}=0$. For finite $T$, the renormalization
stops at $|\epsilon|\sim {T}$.

Beyond the microscopic scale $D_0$, it is instructive to introduce two more
energetic scales that characterize the renormalization of the transmission
coefficient by weak interaction, $D_p$ and $D_r$ (Fig.~\ref{sca}): 
\be \ln
(D_p/D_0)=-1/\alpha~,\quad D_r/D_0={\rm R}_0^{1/2\alpha}~.
\label{4}
\ee
The energy $D_p$ defines the scale on which a perturbation theory in
interaction breaks down. If $|\epsilon|\alt D_p$, the interaction requires a
non-perturbative treatment. The scale $D_p$ does not depend on ${\rm R}_0$ and
is much smaller than $D_0$ for $\alpha\ll 1$. The energy $D_r$ defines the
scale on which a perturbation theory in the impurity strength breaks down. If
$|\epsilon|\alt D_r$, a weak impurity with ${\rm R}_0\ll 1$ yields strong
reflection, ${\rm R}\sim 1$. Provided ${\rm R}_0\ll 1$, the scales $D_p$ and
$D_r$ are parametrically different and, for any $\alpha$, $D_r\ll D_p$. We
will see in Sec.~\ref{III} that the scale $D_r$ is of central importance in RG
theory for a double-barrier structure.

To summarize this section, there are three different types of behavior of
${\rm T}$ with $\epsilon$, as illustrated in Fig.~\ref{sca}:
\begin{itemize}
  
\item[I.]  $D_p\alt |\epsilon|\alt D_0$, a small logarithmic correction
  $\delta {\rm T}\sim \alpha {\rm R}_0 \ln (|\epsilon|/D_0)$ is perturbative
  in both interaction and the impurity strength;
  
\item[II.] $D_r\alt |\epsilon|\alt D_p$, power-law scaling of the correction
  $\delta {\rm T}\sim -{\rm R}_0(|\epsilon|/D_0)^{-2\alpha}$ is perturbative
  in the impurity strength but non-perturbative in interaction;
  
\item[III.]  $|\epsilon|\alt D_r$, power-law vanishing of ${\rm T}\sim
  (|\epsilon|/D_r)^{2\alpha}$ is non-perturbative in both the impurity
  strength and interaction.

\end{itemize}
If the interaction is not weak (i.e., $\alpha$ is not small), the regime I
shrinks to zero while the scaling exponents in the regimes II and III become
$-2\alpha\to 2[(1+2\alpha)^{-1/2}-1]$ and $2\alpha\to 2\alpha_e=
2[(1+2\alpha)^{1/2}-1]$, respectively.\cite{voit94} 

To make a connection with
experiment, it is worth noting that in the Luttinger liquid model ($V_b=0$)
for $N$ channels with the same Fermi velocity and no interchannel transitions,
the constant $\alpha$ is related to the exponent $\alpha_e$ which describes
tunneling into the end of the liquid (see Sec.~\ref{I}) as follows: 
\be
\alpha_e=N^{-1}[\,(1+2N\alpha )^{1/2}-1\,]~. \label{alphaend}
\ee 
E.g., $N=1$
for spinless and $N=2$ for spin-degenerate electrons in a quantum wire with a
single mode of transverse quantization,\cite{voit94} $N=4$ in the
Luttinger-liquid model of the armchair carbon nanotube.\cite{eggergo98,kane97}
Note that $\alpha_e=\alpha$ for any $N$ in the limit of weak interaction
($\alpha\to 0$). For a discussion of the tunneling into the end of a
multi-channel Luttinger liquid with non-equivalent channels see
Ref.~\onlinecite{matveev93a}.

\subsection{Hartree-Fock versus renormalization group}
\label{IIb}

We now turn to a conceptually important point that is highly relevant to our
calculation of the resonant tunneling in Sec.~\ref{III}. The basic idea of
Refs.~\onlinecite{matveev93,yue94} was to relate the RG transformations
(\ref{1}) to the scattering of an electron off Friedel oscillations created by
the impurity screened by other electrons.\cite{meden} Indeed, to first order
in $\alpha$, when the interaction-induced correction to the density
distribution may be neglected, the HF correction to the scattering amplitudes
is logarithmically divergent $\propto \ln (D_0/|\epsilon|)$. The essence of
this singularity is a slow decay of the Friedel oscillations, which behave in
1D, in the absence of interaction, as $|x|^{-1}\sin (2k_F|x|+\phi_F)$. Here
$k_F$ is the Fermi wavevector, $|x|$ the distance to the impurity, and
$\phi_F$ a constant phase shift. This observation suggests that a HF treatment
of the problem would yield a series of leading logarithms $\alpha^n{\cal
  L}^n$. In the HF problem, it would be sufficient to solve iteratively in
$\alpha$ a nonlinear Schroedinger equation
$(H_0+U_{HF}\{\psi\}-\epsilon)\psi=0$, where $H_0$ is the noninteracting
Hamiltonian. The static self-consistent scattering potential is $U_i+U_{HF}$,
where $U_i$ is the bare impurity potential and the nonlocal HF term reads 
\bea
U_{HF}(x,x')=&-&V(x-x')\rho(x,x') \\
&+&\delta (x-x')\!\int\! dx_1 V(x-x_1)\rho(x_1,x_1)~. \nonumber
\label{5}
\eea
Here $V(x)$ is a potential of pairwise interaction between particles,
$\rho(x,x')=\sum_q\psi_q^*(x')\psi_q(x)$ is a sum over all occupied
states, and $\rho(x_1,x_1)$ in the Hartree term is the total electron
density which includes the Friedel oscillations. Self-consistency
requires that both the wave functions $\psi_q$ and the scattering
potential $U_{HF}$ be corrected on every step of the HF iterative
procedure. In effect, Ref.~\onlinecite{yue94} (see also
Ref.~\onlinecite{tsai02}) suggests that the HF procedure, being carried
out in the ``leading-log" approximation, would reproduce
Eq.~(\ref{3}). In particular, the HF calculation\cite{yue94} of second
order in $\alpha$ gives a correction $\sim\alpha^2{\cal L}^2$ and,
according to Ref.~\onlinecite{yue94}, this HF correction does coincide
with the one obtained from an expansion of Eq.~(\ref{3}) up to terms
$\sim {\it O}(\alpha^2)$. Following this logic, the RG (\ref{1}) is
essentially an effective method of solving the HF problem
self-consistently.

In fact, however, while the HF theory is indeed a ``logarithmic
theory" in the sense that it can be expanded in a series of leading
logarithms $\alpha^n{\cal L}^n$, the series is very much different
from the RG solution (\ref{3}), even to one-loop order. To demonstrate
this, let us represent HF solutions for waves $\psi_\pm(x)$ which are
incident on the scatterer from the left (+) and from the right ($-$)
as $\psi_\pm(x)=a_{\pm +}(x)e_+(x)+a_{\pm -}(x)e_-(x)$, where
$e_\pm(x)=\exp [\pm i(k_F+\epsilon/v_F)x]$, $a_{\mu\nu}(x)$ are
amplitudes varying slowly on a scale of $k_F^{-1}$ for $k_F|x|\gg 1$
and satisfying the scattering boundary conditions
$a_{++}(-\infty)=a_{--}(\infty)=1$,
$a_{+-}(-\infty)=a_{-+}(\infty)=0$. Separating the slow and fast
variables, the HF equations 
for $\epsilon_F/|\epsilon|\gg k_F|x|\gg 1$ are written in the
leading-log approximation as 
\bea \partial a_{\pm,+}/\partial {\cal
L}_x&=&(\alpha/2)\,\eta\, a_{\pm,-}~, \nonumber \\ \partial
a_{\pm,-}/\partial {\cal L}_x&=&(\alpha/2)\,\eta^* a_{\pm,+}~,
\label{6} 
\eea 
where ${\cal L}_x=\ln(D_0|x|/v_F)$ and the function
$\eta(x)$ describes the envelope $n(x)=|\eta(x)|/2\pi |x|$ of the Friedel
oscillations. The latter are given for $k_F|x|\gg 1$ by \bea &&{1\over 2\pi
x}\,{\rm Im}\,\{\eta(x)e^{2ik_Fx}\}=\nonumber \\ &&{1\over \pi
v_F}\sum_{\mu=\pm} \int^0_{-D_0}\!\!\!\!d\epsilon\, {\rm
Re}\,\{(a_{\mu +}^*a_{\mu -})e^{-2i(\epsilon_F+\epsilon)x/v_F}\}~.
\label{6a} \eea Solutions of Eqs.~(\ref{6}) for $x>0$ and $x<0$ are
matched onto each other by means of the bare scattering matrix.
Equivalently, Eqs.~(\ref{6}) can be cast in the form 
\be \partial \ln\eta/\partial {\cal
L}_x=\alpha I/2~,\quad \partial I/ \partial {\cal L}_x =2\alpha
|\eta|^2 \label{7} \ee with $\eta =a_{++}a^*_{+-}+a_{-+}a^*_{--}$ and
$I=\sum_{\mu\nu}|a_{\mu\nu}|^2$. The function $I(x)$ is a contribution
of states close to the Fermi energy to a smooth part of electron
density. Equations (\ref{6}),(\ref{7}) can be solved exactly. The
scattering amplitudes and the shape of the Friedel oscillations are
then found self-consistently. However, the rather intricate HF solution
appears to be a qualitatively wrong approximation in our problem of
impurity scattering in an interacting 1D liquid. The difference
between the RG and HF solutions is already seen for a weak impurity
with ${\rm R}_0\ll 1$.  While the RG gives for $|\epsilon|\gg D_r$ a
power-law behavior of the reflection amplitude \be
r_{L,R}=r_{L0,R0}\,\,e^{\alpha {\cal L}}~, \label{8} \ee the HF
expansion yields a geometric progression \be
r_{L,R}=r_{L0,R0}\,\,{1\over 1-\alpha{\cal L}}~.  \label{9} \ee
Equation (\ref{9}) can be obtained also by summing up ladder diagrams
in the particle-hole channel. Note a pole characteristic to HF
approaches.

We thus see that even for weak interaction the HF solution correctly describes
the scattering by a screened impurity only for $|\epsilon|\gg D_p$, where it
gives a small perturbative correction $\sim\alpha{\cal L}$. At the next order
in $\alpha$, the HF expansion does generate a term $\sim
r_{L0,R0}\alpha^2{\cal L}^2$ in $r_{L,R}$ but, in contrast to
Ref.~\onlinecite{yue94}, the numerical coefficient in front of it is a factor
of 2 larger than in Eq.~(\ref{3}), which signifies a breakdown of the HF
approach beyond the simplest Born approximation.

It is worth noting that in Ref.~\onlinecite{tsai02} another HF-type
(``non-selfconsistent HF") scheme was employed to substantiate the RG
procedure.  As mentioned above, single-electron wave functions within the
self-consistent HF approach obey $\psi_q=\psi^0_q+{\hat G}_q^0U_{HF}\psi_q$,
where $\psi^0_q$ and the Green's function operator ${\hat G}^0_q$ describe
noninteracting electrons. Let us write the correction to the transmission
amplitude of second order in $\alpha{\cal L}$ as $\delta t(\epsilon)=-{\cal
  C}t_0{\rm R}_0 (\alpha {\cal L})^2$. The self-consistent HF approximation
gives ${\cal C}={\rm T}_0+1/2$, different from ${\cal C}=(3{\rm T}_0-1)/2$
following from the RG expansion (\ref{3}). In Ref.~\onlinecite{tsai02}, only
$U_{HF}$ is renormalized while $\psi_q$ remain unperturbed, i.e., $\psi_q\to
\psi_q^0$ in the right-hand side of the above HF equation.  This yields ${\cal
  C}= 1$, which accidentally coincides with the RG result for a weak impurity.
However, for arbitrary ${\rm T}_0$ this approach can be seen to fail as well,
already at order ${\it O}(\alpha^2{\cal L}^2)$.

It follows that the RG (\ref{1}) in fact cannot be obtained from the above HF
schemes, i.e., the RG is not a method of solving the HF equations. There are
other scattering processes, not captured by the HF approximation, that at
higher orders in $\alpha$ almost compensate the HF result: a comparison of
Eqs.~(\ref{8}),(\ref{9}) says that the coefficients $c_n$ in the corresponding
sums $\sum_nc_n(\alpha{\cal L})^n$ are different in the RG and HF expansions
by a factor of $n!$.

One can see that the HF approach misses important scattering processes
also by noticing that Friedel oscillations in a Luttinger liquid decay
in the limit of large $|x|\gg v_F/D_r$ as $|x|^{-1+\alpha}$, more
slowly\cite{egger95,egger96} (for the repulsive interaction) than in a
Fermi liquid. The Bragg reflection by a potential created by the
Friedel oscillations can then be easily seen to yield a power-law
behavior of the transmission coefficient only if $\alpha$ is put equal
to zero in the exponent of the Friedel oscillations. For any
$\alpha>0$ the Bragg reflection leads to an exponential decay of the
wave functions, $\ln |\psi|\propto -|x|^\alpha$, instead of a power law
required by self-consistency and by Eq.~(\ref{3}).

We can summarize the above observations by stating that, although it
is certainly appealing to think of the RG (\ref{1}) as being
associated with scattering off Friedel oscillations, the mechanism of
the RG is more complicated.  In fact, in a closely related calculation
of various correlation functions of a clean Luttinger liquid without
impurities the inapplicability of the HF theory was realized long
ago:\cite{bychkov66,dzyaloshinskii72,dzyaloshinskii74} it is the
interaction in a Cooper (particle-particle) channel that interferes
with the HF interaction. As a result, instead of the summation of a HF
ladder one has to use a much more involved parquet
technique.\cite{bychkov66,dzyaloshinskii72,roulet69} In the present
problem, the parquet summation is equivalent to the RG (\ref{1}). A
rigorous microscopic justification of the RG (\ref{1}) can be done by
using Ward identities in a diagrammatic approach and will be given
elsewhere.\cite{inprep} Our purpose in Sec.~\ref{III} is to extend the
RG (\ref{1}) to the case of two impurities.

\section{Double barrier}
\label{III}

Consider now two potential barriers located at $x=0$ and $x=x_0$ and
let the distance $x_0$ be much larger than the width of each of
them. Clearly, the spatial structure itself yields an energy
dependence of the total (describing scattering on both impurities)
amplitudes $t(\epsilon)$ and $r_{L,R}(\epsilon)$, even if such a
dependence may be neglected for each of the impurities, as was assumed
in the derivation of Eqs.~(\ref{1}). Specifically, without interaction
the energy $\Delta=\pi v_F/x_0$ gives a period of oscillations in the
total scattering amplitudes with changing Fermi level. If the
impurities are strong, $\Delta$ is the level spacing inside a quantum
dot formed by the barriers and the period of resonant tunneling
oscillations. It follows that an RG description of scattering off a
double barrier requires a generalization of the
RG~\cite{matveev93,yue94} to the case when the bare amplitudes
$t_0(\epsilon)$ and $r_{L0,R0}(\epsilon)$ are energy dependent.

A question, however, arises if it is at all possible to construct the RG
theory for a compound scatterer in terms of only total $S$-matrix, as in
Eqs.~(\ref{1}). Put another way, the question is if total scattering
amplitudes generated by RG transformations are expressed in terms of
themselves only. As we will see below, the answer depends on the parameter
$\Delta/D_0$. We recall that $D_0$ in our problem is the smallest of two
energy scales given by $\epsilon_F$ and $v_F/d$, where $d$ is the radius of
interaction. If $\Delta\ll D_0$, so that there are many resonances within the
band $(-D_0,D_0)$, the RG transformations generate more terms than are encoded
in the total $S$-matrix. This is a generic case we are interested in. On the
other hand, if $\Delta\gg D_0$ (e.g., when $d\gg x_0$), one would have from
the very beginning a model of a single impurity with no spatial structure but
with possibly energy dependent scattering amplitudes. In that case, it is
sufficient to deal with total amplitudes only (see
Secs.~\ref{IIId},\ref{IIIe}).  For the case of a single resonance this is a
model studied, e.g., in Ref.~\onlinecite{komnik02} (for an exactly solvable
case of the Luttinger liquid parameter $g=1/2$) and
Ref.~\onlinecite{nazarov02} (for a weak interaction, $1-g\ll 1$).

\subsection{Perturbative expansion}
\label{IIIa}

Building on our knowledge of the single impurity case, we start the
derivation of a double barrier RG with a calculation of perturbative
in $\alpha$ corrections to a time-ordered single-particle Green's
function $G_{\mu\nu}({\rm x},{\rm x}')= -i\langle {\cal
T}\psi_\nu({\rm x})\psi^\dagger_\mu({\rm x}')\rangle$, where ${\rm
x}=(x,t)$ and $\mu,\nu = \pm$ label two branches of right (+) and left
($-$) movers. The bare Green's function is characterized by bare
scattering amplitudes $t_0(\epsilon)$, $r_{L0,R0}(\epsilon)$. To first
order in $\alpha$, transforming to the $(x,\epsilon)$-representation,
which is most convenient in the present context, we have a correction
to $G_{\mu\nu}(x_{\rm f},x_{\rm i};\epsilon)$ (summation over branch
indices assumed):
\bea 
\delta G_{\mu\nu}(x_{\rm f},x_{\rm i};\epsilon)=
\int_{-\infty}^{\infty} \!\! dx\, G_{\mu\mu'}(x,x_{\rm i};\epsilon)
\nonumber \\ \times\,\,
\Sigma_{\mu'\nu'}(x)
G_{\nu'\nu}(x_{\rm f},x;\epsilon)~, 
\label{10}
\eea
where 
\be
\Sigma_{\mu\nu}(x)= i\alpha \, v_F\!\int_{-D_0}^{D_0} \!\!\!
d\epsilon\, G_{\mu\nu}(x,x;\epsilon)~.
\label{11} 
\ee 
In the Luttinger liquid model, only forward scattering due to interaction is
present: we explicitly assume this in Eqs.~(\ref{10}),(\ref{11}) by writing
the self-energy that depends on two branch indices only [backward scattering
can be straightforwardly incorporated for spinless electrons, see
Eq.~(\ref{2})]. As explained in Sec.~\ref{IIa}, we also assume in
Eq.~(\ref{10}) that the interaction is effectively short-ranged and write the
self-energy as a spatially local quantity. Indeed, the perturbative
logarithmic correction to $G_{\mu\nu}(x_{\rm f},x_{\rm i};\epsilon)$
comes from energies $|\epsilon|\alt v_F/d$, i.e., from spatial scales $|x|$
larger than $d$. We start with the case $d\ll x_0$ (we will return to the
simpler case $d\gg x_0$ in Sec.~\ref{IIId}). Moreover, in the following, we
neglect the forward scattering of electrons belonging to the same branch: to
one-loop order, such processes can be seen to yield only a non-singular
renormalization of the bare parameters. The forward-scattering interaction
that generates RG transformations similar to Eqs.~(\ref{1}) is that of
electrons from different branches.\cite{g-ology} We thus formulate an impurity
problem with only non-diagonal couplings $\Sigma_{+-}(x)$ and
$\Sigma_{-+}(x)$.

The scattering amplitudes are related to $G_{\mu\nu}(x_{\rm f},x_{\rm
i};\epsilon)$ for $\epsilon>0$ (we measure $\epsilon$ from the Fermi
level upwards) as
\bea
r_L(\epsilon)&=&i v_F G_{+-}(x_{\rm f}, x_{\rm i};\epsilon)
e^{-i\epsilon(x_{\rm f}+x_{\rm i})/v_F}|_{x_{\rm f},x_{\rm i}\to
-\infty}~, \nonumber \\ 
r_R(\epsilon)&=&i v_F G_{-+}(x_{\rm f}, x_{\rm i};\epsilon)
e^{i\epsilon(x_{\rm f}-x_{\rm i})/v_F}|_{x_{\rm f},x_{\rm i}\to
\infty}~, \label{12}\\ 
t(\epsilon)&=&i v_F G_{++}(x_{\rm f}, x_{\rm i};\epsilon)
e^{-i\epsilon(x_{\rm f}-x_{\rm i})/v_F}|_{x_{\rm f}\to \infty,x_{\rm i}\to
-\infty}~, \nonumber
\eea
and similarly for $\epsilon<0$ by changing $\pm \to \mp$ in branch
indices of $G_{\mu\nu}$. We count the phases of the reflection
amplitudes from $x=0$. Since the integral over $x$ in Eq.~(\ref{10})
involves integration over the interval of $x$ inside the dot,
$0<x<x_0$, corrections $\delta t(\epsilon)$, $\delta
r_{L,R}(\epsilon)$ cannot be expressed solely in terms of the bare
amplitudes $t_0(\epsilon)$ and $r_{L0,R0}(\epsilon)$. A closed set of
equations can be written by introducing amplitudes to stay
inside the dot
\be
A_{\mu,-\mu}(\epsilon)=iv_F G_{\mu,-\mu}(x,x;\epsilon)
e^{ 2i\mu\epsilon x} 
\label{13} 
\ee
(we need only nondiagonal amplitudes $A_{\mu,-\mu}$), amplitudes to
escape from the dot, to the left or to the right,
\be
d_{\mu}^{\pm}(\epsilon)=iv_F G_{\mu,\pm}(x_{\rm f},x ;\epsilon)
e^{-i\epsilon(x_{\rm f}-\mu x)}|_{x_{\rm f}\to\pm\infty}~, 
\label{14}
\ee 
and, similarly, amplitudes to get into the dot from outside
\be
b_{\mu}^{\pm}(\epsilon)= iv_F G_{\pm,\mu}(x,x_{\rm i};\epsilon)
e^{i\epsilon(x_{\rm i}\mp x)}|_{x_{\rm i}\to \mp\infty}~. 
\label{15}
\ee
The amplitudes (\ref{12})--(\ref{15}) are constrained by unitarity and,
moreover, $b_\mu^\nu(\epsilon)=d_{-\mu}^\nu(\epsilon)$
by time reversal symmetry which we assume throughout the paper. In
Eqs.~(\ref{13})--(\ref{15}), $x$ lies within the dot. Without
interaction, the amplitudes (\ref{13})--(\ref{15}) do not depend on $x$.
This property is preserved in the leading-log approximation.

The first-order corrections to $t(\epsilon)$ and $r_L(\epsilon)$ read
\bea
\delta t(\epsilon) = -{\alpha \over 2}\int_{-D_0}^{D_0}
\frac{d\epsilon'}{\epsilon-\epsilon'} {\Big\{
}L_{+}(\epsilon,\epsilon') \qquad\qquad\qquad \label{16} \\
+\theta(-\epsilon')t(\epsilon)
[r_R(\epsilon)r^*_R(\epsilon')\chi_{\epsilon-{\epsilon'}}
+r_L(\epsilon) r^{*}_L(\epsilon')] {\Big\} }~, \nonumber \\ \delta
r_L(\epsilon) = -{\alpha \over 2} \int_{-D_0}^{D_0}
\frac{d\epsilon'}{\epsilon-\epsilon'} {\Big\{ }
L_{-}(\epsilon,\epsilon')+ \theta(\epsilon')r_L(\epsilon') \qquad
\label{17} \\ + \theta(-\epsilon')[ t^2(\epsilon)
r^*_R(\epsilon')\chi_{\epsilon-{\epsilon'}} +r_L^2(\epsilon)
r^{*}_L(\epsilon')] {\Big\}}~,
\nonumber
\eea
and similar equations can be written for corrections to
$r_R(\epsilon)$ and the amplitudes defined in
Eqs.~(\ref{13})--(\ref{15}).  Here $\theta(\epsilon)$ is the step
function and 
\be
\chi_\epsilon=\exp(2\pi i\epsilon/\Delta)~.
\label{17a}
\ee
The terms
$L_\mu(\epsilon,\epsilon')$, which correspond to the integration over
$x$ in (\ref{10}) inside the dot, are given by
\bea
L_{\mu}(\epsilon,\epsilon')&=&
b_{+}^{-}(\epsilon)A_{+-}(\epsilon')d_{-}^{\mu}(\epsilon)
(\chi_{\epsilon-\epsilon'}-1)\nonumber \\
&+&b_{-}^{-}(\epsilon)A_{-+}(\epsilon')d_{+}^{\mu}(\epsilon)
(1-\chi_{\epsilon'-\epsilon})~.
\label{18}
\eea
The amplitudes to stay inside the dot satisfy the relation
$A_{\mu,-\mu}(\epsilon)=-A^*_{-\mu,\mu}(-\epsilon)$ and it is useful to
decompose them as
\be
A_{\mu,-\mu}(\epsilon)=\theta(\epsilon)B_{\mu,-\mu}
(\epsilon)+\theta(-\epsilon)C_{\mu,-\mu}(\epsilon)~.
\label{18a}
\ee
We thus arrive at a closed system of perturbative equations written in
terms of quantities describing the dot as a whole, without directly
referring to parameters characterizing two barriers.

\subsection{Renormalization group for a double barrier}
\label{IIIb}

Generically (singular exceptions are discussed below), the above
first-order corrections diverge logarithmically as $\epsilon\to 0$,
which implies that higher orders of the perturbation theory are
important.  Naively, one could try to treat Eqs.~(\ref{16}),(\ref{17})
and other perturbative equations self-consistently, i.e., not as
expressions for small corrections $\delta t(\epsilon)$, etc., but as
equations to be solved for $t(\epsilon)=t_0(\epsilon)+\delta
t(\epsilon)$, etc., where $\delta t(\epsilon)$ is not necessarily
small. However, this would not correctly describe the case of strong
impurities, either the case of a dot formed by two barriers or even
that of a single strong barrier. To see this in the simpler case of a
single impurity, take the limit $x_0\to 0$. The terms
$L_\mu(\epsilon,\epsilon')$ and factors $\chi_{\epsilon-\epsilon'}$
then drop out in Eqs.~(\ref{16}),(\ref{17}) and the self-consistent
equations acquire the form
\bea
t(\epsilon)&=&t_0(\epsilon)\label{19} \\ &+& {\alpha \over 2}
\int_{-D_0}^{-|\epsilon|} \!\!\frac{d\epsilon'}{\epsilon'}
t(\epsilon)[\,r_R(\epsilon)r^*_R(\epsilon') + r_L(\epsilon)
r^{*}_L(\epsilon')\,]~,
\nonumber
\eea \vspace{-0.5cm}
\bea
r_L(\epsilon) &=&r_{L0}(\epsilon) + {\alpha \over 2}
\int_{|\epsilon|}^{D_0} \!\!{d\epsilon'\over \epsilon'}r_L(\epsilon')
\label{20} \\ &+& {\alpha\over 2}\int_{-D_0}^{-|\epsilon|}
\!\!{\frac{d\epsilon'}{\epsilon'}}[\,t^2(\epsilon) r^*_R(\epsilon') +
r_L^2(\epsilon) r^{*}_L(\epsilon')\,]~.
\nonumber
\eea
As a comparison of Eqs.~(\ref{16}),(\ref{17}) and (\ref{19}),(\ref{20}) shows,
the latter do not describe two impurities since they miss the terms
$L_\mu(\epsilon,\epsilon')$ and the factors $\chi_{\epsilon-\epsilon'}$.
Moreover, Eqs.~(\ref{19}),(\ref{20}) do not describe even a single strong
structureless (with no dependence of the bare amplitudes on $\epsilon$)
impurity. This can be checked
straightforwardly, e.g., by putting all reflection coefficients in the
integrand of Eq.~(\ref{19}) equal to unity: in this limit the integration over
$\epsilon'$ in Eq.~(\ref{19}) gives a geometric progression of logarithms,
instead of reproducing Eq.~(\ref{1}) for $t(\epsilon)$. Also, plugging in
Eq.~(\ref{3}) into the integrand of Eq.~(\ref{20}) can be easily seen to give
a logarithmic divergence at $\epsilon\to 0$, instead of a power law. The
logarithmic divergencies are characteristic to HF approaches discussed in
Sec.~\ref{IIb}. Equations (\ref{19}),(\ref{20}), however, do describe
correctly a weak structureless impurity, but only for $|\epsilon|\gg D_r$. 

We now derive nonperturbative amplitudes for a double barrier using an
appropriate RG scheme. To account for the $\epsilon$ dependence of the bare
amplitudes, the derivation of the RG from the perturbative results
(\ref{16}),(\ref{17}) necessitates introduction of two energies, $\epsilon$
and $D$. The latter is a flow parameter in RG transformations, i.e., an
ultraviolet cutoff rescaled after tracing over states with energies
$\epsilon'$ in the interval $|\epsilon'|\in (D,D_0)$. The renormalization
stops at $D={\rm max}\{|\epsilon|, T\}$. 

The essence of the RG procedure\cite{matveev93,yue94} is a perturbative
treatment of contributions to the renormalized amplitudes at energy $\epsilon$
from all states with energies $\epsilon'$ in the interval $|\epsilon'|\in
(D,\Lambda D)$, starting from $D=D_0/\Lambda$, such that $\Lambda \gg 1$ but
$\alpha\ln\Lambda \ll 1$.\cite{without} The RG equations thus differ from both
the HF equations and Eqs.~(\ref{19}),(\ref{20}) in that all the amplitudes
depend on $D$ and, moreover, the HF-type integration over projected states
with energies $\epsilon'$ only goes over the interval $|\epsilon'|\in
(D,\Lambda D)$ instead of $(0,D_0)$.

In effect, each step of the RG transformations accounts for the scattering off
the Friedel oscillations in a {\it finite} spatial region, $|x|,|x-x_0|\in
(v_F/\Lambda D, v_F/D)$. Moreover, the Friedel oscillations are only partly
modified, through the (already performed) renormalization of the reflection
amplitudes at energies larger than $D$. At the same time, the scattering
matrix at energies smaller than $D$ is taken at its bare value. This should be
contrasted with the HF approach, where the scattering amplitudes are
determined by interaction processes on all energy scales on every step of the
HF iterations.

The system of one-loop RG equations for a double barrier reads
\bea
\frac{\partial t(\epsilon,D)}{\partial {\cal L}_D} 
&=&{\hat I}_{\epsilon'}(\epsilon,D)  
{\Big\{}  L_{+}(\epsilon,\epsilon';D)\nonumber \\
&+& \theta(-\epsilon')t(\epsilon,D) 
[\,r_R(\epsilon,D)r^*_R(\epsilon',D)\,\chi_{\epsilon-{\epsilon'}}
\nonumber \\
&+&r_L(\epsilon,D)r^{*}_L(\epsilon',D)\,] 
{\Big\} }~,  
\label{21} 
\eea
\bea
\frac{\partial r_L(\epsilon,D)}{\partial {\cal L}_D}  
&=& {\hat I}_{\epsilon'}(\epsilon,D) \
{\Big\{ } L_{-}(\epsilon,\epsilon';D)+
\theta(\epsilon')r_L(\epsilon',D) \nonumber
\\ 
&+&\theta(-\epsilon') 
[\,t^2(\epsilon,D) r^*_R(\epsilon',D) \,\chi_{\epsilon-{\epsilon'}}
\nonumber \\&+& r^2_L(\epsilon,D) r^{*}_L(\epsilon',D)\,] 
{\Big\} }~,  
\label{22}
\eea
and similar equations for other amplitudes.  Here ${\cal L}_D=\ln
(D_0/D)$ and we introduced $D$ dependent amplitudes $t(\epsilon,D)$,
etc., so that integration of Eqs.~(\ref{21}),(\ref{22}) over $D$ acts
on all the amplitudes [which should be contrasted with
Eqs.~(\ref{19}),(\ref{20}), where the integration acts on only one
amplitude in the products of three]. All amplitudes in
$L_\mu(\epsilon,\epsilon')$ (\ref{18}) are now also functions of
$D$. The integral operator ${\hat I}_{\epsilon'}(\epsilon,D)$ is
defined as
\be
{\hat I}_{\epsilon'}(\epsilon,D)=
-\frac{\alpha}{2\ln\Lambda}
\left[\int_D^{\Lambda D} \!\!\!+ 
\int_{-\Lambda D}^{- D}\right]
\frac{d\epsilon'}{\epsilon-\epsilon'}
\left\{\ldots\right\}~,
\label{23}
\ee
where $\Lambda\gg 1$ is restricted by the condition
$\alpha\ln\Lambda\ll 1$. The theory is renormalizable if the action of
the operator ${\hat I}_{\epsilon'}(\epsilon,D)$ to leading order is
independent of $\Lambda$, which is the case for the present problem
within the leading-log approximation. Needless to say, in the limit $x_0\to 0$ 
[i.e., $L_\pm(\epsilon,\epsilon';D)\to 0$ and $\chi_{\epsilon-\epsilon'}\to
1$] Eqs.~(\ref{21}),(\ref{22}) describe a single impurity. 

At finite temperature $T$, one should substitute the Fermi
distribution function $n_F(\epsilon)$ for the step functions in
Eqs.~(\ref{21}),(\ref{22}) and also in Eq.~(\ref{18a}) according to
$\theta(\pm\epsilon)\to n_F(\mp\epsilon)$. The factor
$(\epsilon-\epsilon')^{-1}$ in Eq.~(\ref{23}) effectively stops the
renormalization at $D\sim |\epsilon|$, while the factors
$n_F(\pm\epsilon')$ do so at $D\sim {T}$, otherwise the
renormalization can be carried out down to $D=0$. The infrared cutoff
at $D\sim {T}$ establishes a characteristic spatial scale of
$L_{T}=v_F/{T}$. Due to the thermal smearing, the Friedel
oscillations decay exponentially on a scale of $L_{T}$ [which can be
seen from Eq.~(\ref{6a}) if one plugs in $n_F(\epsilon)$ into the
integrand and extends the integration to $\epsilon>0$]. The absence of
renormalization at $D\ll {T}$ is closely related to the thermal
cutoff of interaction-induced corrections\cite{altshuler85} to the
conductance in higher dimensions.

The RG equations (\ref{21}),(\ref{22}) should be solved with proper
boundary conditions at $D=D_0$, where the renormalization starts from
the bare values of the amplitudes (\ref{12})--(\ref{15}). For the
double-barrier problem, the boundary conditions for the transmission
and reflection amplitudes are:
\be
t_0(\epsilon)={t_1t_2\over S(\epsilon)}~, \quad
r_{L0}(\epsilon)=r_1+{r_2t_1^2 \over S(\epsilon)}\,\chi_\epsilon~, 
\label{24}
\ee
where 
\be
S(\epsilon)=1-r_2r_1'\chi_\epsilon~,
\label{24a}
\ee
and 
$r_{R0}(\epsilon)=-r_{L0}^*(\epsilon)t_0(\epsilon)/
t_0^*(\epsilon)$ by unitarity.
The coefficients $r_{1,2} (r_{1,2}')$ are the noninteracting reflection
amplitudes from the left (right) and $t_{1,2}$ are the transmission
amplitudes of each of the two barriers, respectively. Similarly for
other amplitudes:
\bea
d^{+}_{+}(\epsilon)&=&
{d^{+}_{-}(\epsilon)\over r_1'}={t_0(\epsilon) \over t_1}~, \nonumber\\
d^{-}_{-}(\epsilon)&=&{d^{-}_{+}(\epsilon)\over r_2\,\chi_\epsilon}=
{t_0(\epsilon) \over t_2}~, \label{25}\\
B_{+-}(\epsilon)&=&{t_0(\epsilon) r_2 \over
t_1t_2}\,\chi_\epsilon~,\quad C_{+-}(\epsilon)=-\left[{t_0(\epsilon)
r_1'\over t_1t_2}\right]^*. \nonumber
\eea
We count the phases of $r_{L,R}$ from $x=0$ and the phases of $r_{1,2}$
are defined for an impurity sitting at $x=0$.

\subsection{Separate renormalization of two \\ impurities: $D\gg\Delta$}
\label{IIIc}

We are now in a position to solve the system of RG equations
(\ref{21}),(\ref{22}) by integrating out all states with energies
$|\epsilon'|\agt \max\{|\epsilon|,{T}\}$. We begin with the case
$D_0\gg \Delta$, which is a typical case unless interaction is very
long ranged. We proceed in two steps. Let us first integrate over
$D\gg \Delta$. This can be done for arbitrary $\epsilon$.
Specifically, if $|\epsilon|\agt\Delta$, this will already solve the
problem by providing us with fully renormalized amplitudes. In the
more interesting case of $|\epsilon|\alt\Delta$, we will only sum up
contributions to the renormalized amplitudes from states with
$|\epsilon'|\agt\Delta$ and, as a second step, will have to proceed
with renormalization for $D\alt\Delta$.

Since the renormalization for $D\gg\Delta$ involves many resonant
levels, the amplitudes contain slowly varying parts and parts
oscillating rapidly with changing $\epsilon'$ on a scale of
$\Delta$. Integration over $\epsilon'$ in Eq.~(\ref{23}) allows us to
separate the slow and fast variables: as a result, the dependence of
the amplitudes on $D$ will be slow on the scale of $\Delta$.
However, even after that the RG equations look rather
cumbersome. To construct the solution to these equations, note that an
important parameter $D/D_{r_{\rm min}}$ is available, where 
\be
D_{r_{\rm
min}}=\min \{D_{r_1},D_{r_2}\}
\label{25a-1}
\ee
and $D_{r_{1,2}}$ are defined for each
of two barriers by Eq.~(\ref{4}). If both barriers are initially
(i.e., at $D=D_0$) strong ($|t_{1,2}|\ll 1$), then this parameter is
small for all $D<D_0$. However, if one or both of the barriers are
initially weak, there is a range of $D\in (D_{r_{\rm min}},D_0)$ where
at least one barrier still remains weak.

\begin{figure}[ht]
\includegraphics[width=8cm]{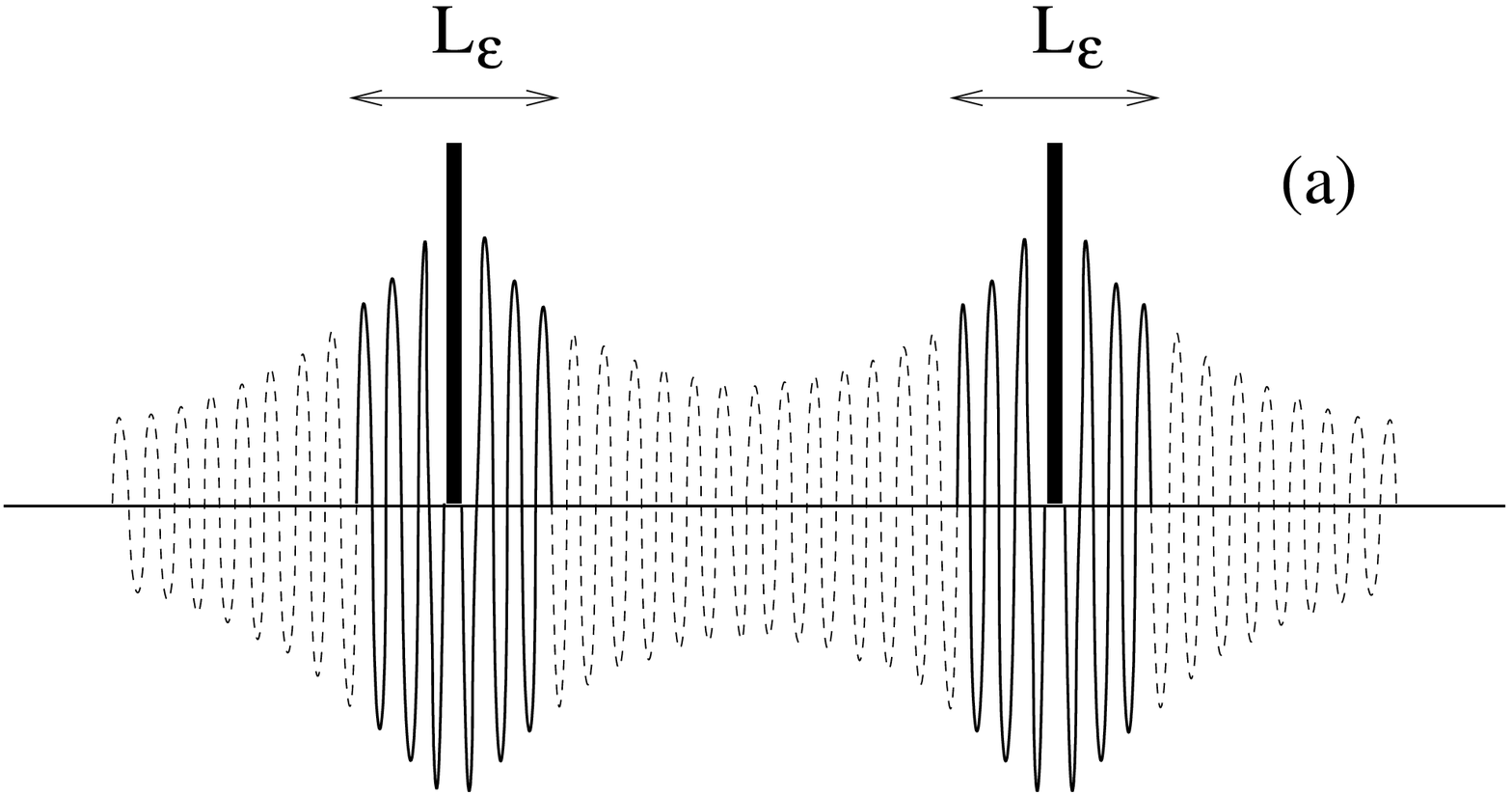}

\vspace{5mm}

\includegraphics[width=8cm]{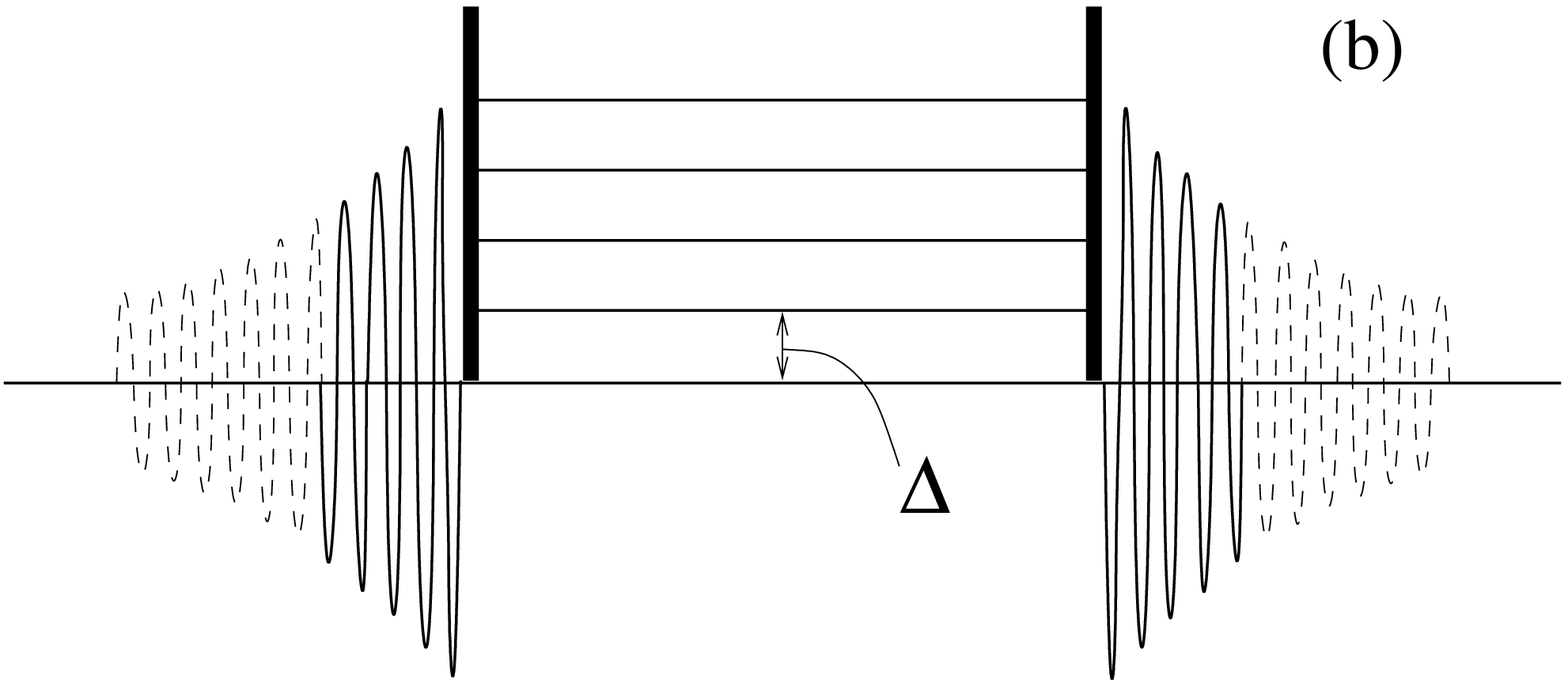}
\caption{\label{osc} Sketch of different stages of the RG
  procedure for a double barrier for $D\gg \Delta \gg T$,
  Eqs.~(\ref{28})-(\ref{29a}).  The Friedel oscillations shown schematically
  around each of the barriers yield a renormalization of the scattering
  amplitudes and are renormalized themselves as well. At energy $\epsilon$,
  the renormalization comes from spatial scales smaller than
  $L_\epsilon=v_F/|\epsilon|$ (these regions are marked by the solid lines).
  (a) Separate renormalization of two {\it weak} barriers.  Since there is no
  energy quantization between the barriers, the Friedel oscillations both
  inside and outside the dot contribute to the renormalization and the
  barriers ``do not talk to each other". The scaling exponent $\alpha$ is the
  same as for a single barrier, Sec.~\ref{IIa}. (b) Separate renormalization
  of two {\it strong} barriers. This figure describes either initially strong
  barriers or those that are initially weak but both become strongly
  reflecting due to the renormalization. The inner part of the dot with a
  discrete energy spectrum does not contribute to the renormalization. The
  renormalization of each of the barriers is governed by the exponent
  $\alpha/2$, twice as small as for weak barriers.}
\end{figure}

It is useful first to examine some general properties of integrals
over $\epsilon'$ that appear in the course of renormalization. Let us
return to the perturbative expansion (\ref{16}),(\ref{17}). We see that the
averaging over $\epsilon'$ involves two types of integrals
\be
{\cal I}_1=\int_{|\epsilon|}^{D_0}\!\!{d\epsilon'\over
\epsilon'}{1\over S(\epsilon')}~,\quad {\cal I}_2=
\int_{|\epsilon|}^{D_0}\!\!{d\epsilon'\over
\epsilon'}{\chi_{\epsilon'}\over S(\epsilon')}~,
\label{25a}
\ee
where $S(\epsilon)$ is given by Eq.~(\ref{24a}) and ${\cal I}_{1,2}$ are
related by ${\cal I}_2=({\cal I}_1-{\cal L})/r'_1r_2$. The integrals
(\ref{25a}) are evaluated in different ways depending on whether at least one
of the barriers is weak (so that $|r'_1r_2|\ll 1$) or both barriers are strong
($|r'_1r_2|\simeq 1$). In the former case the integrand of ${\cal I}_1$ is
only slightly modulated, so that one can expand the factor $S^{-1}(\epsilon')$
and average over harmonics $\chi^n(\epsilon')$. Then only zero harmonics give
rise to singular (logarithmic) corrections and in the leading-log
approximation we have
\be
{\cal I}_1={\cal L}~,\quad {\cal I}_2=0~.
\label{25b}
\ee 
In the opposite case of strong barriers, sharp resonances appear that
are described by a Breit-Wigner formula for $S^{-1}(\epsilon')$ and
yield $|{\cal I}_1|\simeq |{\cal I}_2|$:
\be
{\cal I}_1={\cal L}/2~,\quad {\cal I}_2=-{\cal L}/2r'_1r_2~.
\label{25c}
\ee 
The situation repeats itself in the RG equations (\ref{21}),(\ref{22}).  The
difference in the factor of 1/2 between the values of ${\cal I}_1$ in
Eqs.~(\ref{25b}),(\ref{25c}) implies that the renormalization should be
carried out differently (see Fig.~\ref{osc}) in the regions $D\gg D_{r_{\rm
    min}}$ and $\Delta\ll D\ll D_{r_{\rm min}}$.

For $D\gg D_{r_{\rm min}}$, similarly to Eq.~(\ref{25b}), after the averaging
over $\epsilon'$ only zero harmonics contribute to the renormalization. The
solution at $D\gg D_{r_{\rm min}}$ is then simple: it has the form of
Eqs.~(\ref{24}),(\ref{25}) with the reflection and transmission amplitudes of
each of two barriers renormalized separately (Fig.~\ref{osc}a), according to
the RG (\ref{1}),(\ref{3}) for a single impurity: 
\bea 
\frac{\partial
  t_{1,2}(D)}{\partial {\cal L}_D} &=&
-\alpha t_{1,2}(D)|r_{1,2}(D)|^2~, \label{26} \\
\frac{\partial r_{1,2}(D)}{\partial {\cal L}_D} &=& \ \alpha
r_{1,2}(D)|t_{1,2}(D)|^2~.
\label{27}
\eea
One can check straightforwardly that Eqs.~(\ref{24}),(\ref{25})
(describing the Fabry-Perot resonance) with the replacement
\bea
t_{1,2}&\to& \frac{(D/D_0)^\alpha
t_{1,2}}{[\,|r_{1,2}|^2+(D/D_0)^{2\alpha}|t_{1,2}|^2\,]^{1/2}}~,
\label{28} \\ r_{1,2}(r_{1,2}')&\to&
\frac{r_{1,2}(r_{1,2}')}{[\,|r_{1,2}|^2+
(D/D_0)^{2\alpha}|t_{1,2}|^2\,]^{1/2}}
\label{29}
\eea
solve Eqs.~(\ref{21}),(\ref{22}) averaged over harmonics for $D\gg
\max \{ D_{r_{\rm min}},\Delta\}$. 

On the other hand, if $D_{r_{\rm min}}\gg\Delta$, there is an interval
of $D\in (\Delta, D_{r_{\rm min}})$ in which each of the impurities is
strongly reflecting (Fig.~\ref{osc}b), so that the averaged equations
are again simplified by summing over resonance poles, similarly to
Eq.~(\ref{25c}). We get an independent renormalization of
$t_{1,2}(D)$ according to
\be
t_{1,2}(D)/t_{1,2}(D_{r_{\rm min}})=(D/D_{r_{\rm min}})^{\alpha/2}~,
\label{29a}
\ee
and the scaling exponent is now half that for $D\gg D_{r_{\rm min}}$. 
We thus have two solutions given by Eqs.~(\ref{28}),(\ref{29}) and
Eq.~(\ref{29a}), respectively, that match onto each other at $D\sim
D_{r_{\rm min}}$. Due to the slow power-law dependence, for $\alpha\ll
1$ the matching is exact.

We conclude that the key difference between the renormalization for $D$ larger
and smaller than $D_{r_{\rm min}}$ is that for $D\gg D_{r_{\rm min}}$ the
transmission amplitudes for each barrier are renormalized with the exponent
$\alpha$, whereas for $D \ll D_{r_{\rm min}}$ with the exponent $\alpha/2$. In
both limiting cases, two barriers are renormalized separately for $D\gg\Delta$
(Fig.~\ref{osc}). It is worth stressing that generally the independent
renormalizations of two barriers cannot be derived from RG equations written
in terms of only $t(\epsilon)$ and $r_{L,R}(\epsilon)$, i.e., the terms
$L_\mu(\epsilon,\epsilon';D)$ are of crucial importance in the derivation of
Eqs.~(\ref{26})--(\ref{29}). However, the renormalization of resonant
tunneling amplitudes for energies near resonances allows for another
formulation which involves $t(\epsilon)$ and $r_{L,R}(\epsilon)$ only, we will
return to this issue in Sec.~\ref{IIId}.

If $\epsilon$ is close to one of resonant energies,
Eqs.~(\ref{21}),(\ref{22}) can be further simplified by expanding
$\chi_\epsilon$ near the resonance: the renormalized amplitudes for
strong barriers take then the form of Breit-Wigner amplitudes with $D$
dependent widths $\Gamma_{1,2}(D)=(\Delta/2\pi)|t_{1,2}(D)|^2\propto
D^\alpha$. Specifically, for initially strong barriers:
\be
\Gamma_{1,2}(D)={\Delta\over 2\pi}\,|t_{1,2}|^2\left({D\over
D_0}\right)^\alpha,
\label{29b}
\ee
where $|t_{1,2}|^2\ll 1$ are the bare transmission probabilities at
$D=D_0$ and resonant peaks are sharp [i.e., $\Gamma_{1,2}(D)\ll
\Delta$] for all $D<D_0$. If at least one
barrier is initially weak, the resonant structure develops only at
$D\ll D_{r_{\rm min}}$. Provided  one barrier
is initially weak (assume this is the right barrier and $D_{r_2}\gg
\Delta$), whereas the other is strong, then 
\bea
\Gamma_1(D)&=&{\Delta\over 2\pi}\,|t_1|^2\left({DD_{r_2}\over
D_0^2}\right)^{\alpha},
\label{29c} \\
\Gamma_2(D)&=&{\Delta\over
2\pi}\left({D\over D_{r_2}}\right)^\alpha.\label{29d}
\eea 
If $\Delta\ll D_{r_2}\alt
D_{r_1}\ll D_0$, i.e., both barriers are initially weak, then
\be
\Gamma_1(D)={\Delta\over 2\pi}\left({DD_{r_2}\over
D_{r_1}^2}\right)^{\alpha}
\label{29e}
\ee
and $\Gamma_2(D)$ is given again by Eq.~(\ref{29d}).

To summarize this section, substituting $D\to |\epsilon|$ in
Eqs.~(\ref{28}),(\ref{29}),(\ref{29a}) and using the Fabry-Perot
equations (\ref{24}) gives the fully renormalized scattering
amplitudes for $|\epsilon|\agt \Delta$ if $|\epsilon|\gg 
T$. Also, if ${T}\gg \Delta$, substituting $D\to {T}$ solves
the problem for arbitrary $\epsilon$. However, when both
$|\epsilon|,{T}\ll \Delta$, we should proceed with the
renormalization in the range $D\ll\Delta$.

\subsection{Single resonance: $D\ll\Delta$}
\label{IIId} 

\begin{figure}[ht]
\includegraphics[width=8cm]{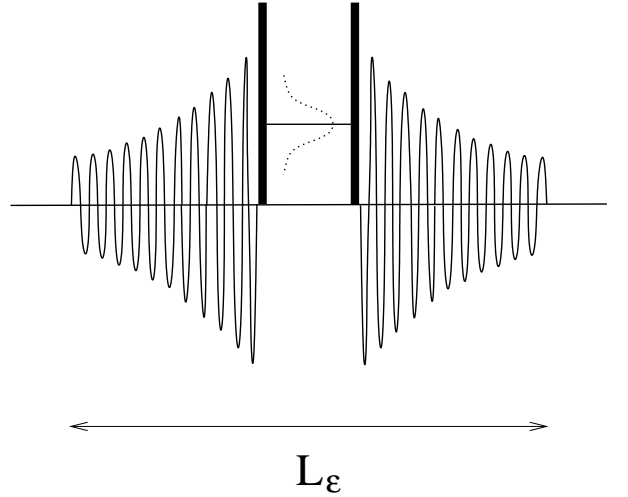}
\caption{\label{osc1} Illustration of the renormalization
  of a double barrier for $D\ll \Delta$, Eqs.~(\ref{30})-(\ref{37a}).  The
  Friedel oscillations are shown schematically within the range
  $L_\epsilon=v_F/|\epsilon|$ outside the dot, where they contribute to the
  renormalization of the scattering amplitudes at energy $\epsilon$ (provided
  $|\epsilon|\gg T$). The double barrier for small $D$ may be considered as a
  single barrier with an energy dependent scattering matrix, which describes
  the resonance (sketched by the dotted line) on a single level. The
  scattering matrix inside the renormalized resonance is discussed in
  Sec.~\ref{IIIe}. }
\end{figure}

Let us now consider Eqs.~(\ref{21}),(\ref{22}) for
$|\epsilon|,|\epsilon'|\ll\Delta$. In this limit, the terms (\ref{18})
containing the amplitudes $A_{\mu,-\mu}(\epsilon')$ to stay inside the dot
become irrelevant in the RG sense: the phase factors
$\chi_{\epsilon-\epsilon'}$ in Eq.~(\ref{18}) can be expanded about
$\epsilon,\epsilon'=0$, which leads to the cancellation of the singular factor
$(\epsilon-\epsilon')^{-1}$ in Eq.~(\ref{23}). As a result, the terms
$L_\mu(\epsilon,\epsilon';D)$ do not contribute to the renormalization at
$D\ll\Delta$.  The factors $\chi_{\epsilon-\epsilon'}$ should also be omitted
in the terms of Eqs.~(\ref{21}),(\ref{22}) that are proportional to
$r_R^*(\epsilon')$. Thus we are led to a coupled set of RG equations that
describe also a single impurity with energy dependent scattering amplitudes:
the spatial structure of the double barrier system is of no importance for the
renormalization at $D\ll\Delta$ (see Fig.~\ref{osc1}). However, the boundary
conditions in the double-barrier case should be written at $D\sim \Delta$,
instead of $D \sim D_0$. We obtain for $|\epsilon|,|\epsilon'|,D\ll\Delta$:
\bea 
\frac{\partial t(\epsilon,D)}{\partial {\cal L}_D} &=& -{\alpha \over 2}
\ t(\epsilon,D)
\,[\, r_R(\epsilon,D)\ \overline{r_R^*}(D)\nonumber \\
&&+r_L(\epsilon,D)\ \overline{r_L^*}(D) \,]~, \label{30} \\
\frac{\partial r_L(\epsilon,D)}{\partial {\cal L}_D} &=& {\alpha \over 2}
\,[\,\overline{r_L}(D)
-t^2(\epsilon,D)\ \overline{r_R^*}(D) \nonumber \\
&& -r_L^2(\epsilon,D)\ \overline{r_L^*}(D) \,]~,
\label{31}
\eea
and similarly for $r_R(\epsilon,D)$, where the bar over the reflection
amplitudes denotes the averaging (\ref{23}) over $\epsilon'$.

We integrate now Eqs.~(\ref{30}),(\ref{31}) assuming that each of two
barriers is characterized by $D_{r_{1,2}}\gg\Delta$. This condition
means that either the barriers are strong initially at $D=D_0$ or get
strong in the course of renormalization (\ref{26}),(\ref{27}) before
$D$ equals $\Delta$. We will analyze the case of both or one of
$D_{r_{1,2}}$ being smaller than $\Delta$ in Sec.~\ref{IIIf}.

Thus, if $D_{r_{1,2}}\gg\Delta$, we have narrow peaks of resonant
transmission in a neighborhood of the Fermi level, even if the bare
reflection coefficients are small. Consider first the case of a
resonance energy $\epsilon_0$ lying exactly on the Fermi level,
$\epsilon_0=0$. Let $\Gamma(D)$ be a renormalized width of the
resonance peak at the Fermi energy (to be found below). If $D\gg
\Gamma(D)$, then $|\overline{r_{L,R}}(D)|\simeq 1$, which allows for a
significant simplification of Eqs.~(\ref{30}),(\ref{31}). It is
convenient to introduce phase-shifted amplitudes ${\tilde
r}_L=r_Le^{-i\varphi_{r_1}}$, ${\tilde
r}_R=r_Re^{-i\varphi_{r'_2}+2\pi i(\epsilon_F+\epsilon_0)/\Delta}$,
${\tilde t}=te^{-i(\varphi_{t_1}+\varphi_{t_2})}$, where
$\varphi_{r_1}$ is the phase of $r_1$, etc., in obvious notation. Then
we get, by putting the averaged amplitudes far from the resonance
$\overline{{\tilde r}_{L,R}}(D) =1$:
\bea
&&\frac{\partial \tilde{t}(\epsilon,D)}{\partial {\cal L}_D} =
-{\alpha \over 2} \ \tilde{t}(\epsilon,D) \,[\,
\tilde{r}_L(\epsilon,D)+\tilde{r}_R(\epsilon,D)\, ]~, \ \ \label{32}\\
&&\frac{\partial \tilde{r}_{L,R}(\epsilon,D)}{\partial {\cal L}_D} =
{\alpha \over 2} \,[\,
1-\tilde{r}_{L,R}^2(\epsilon,D)-\tilde{t}^2(\epsilon,D)\, ]~,
\label{33}
\eea
with the following solutions 
\bea
&&\tilde{t}(\epsilon,D)=\frac{[u_{+}^2(D)-u_{-}^2(D)]^{1/2}}
{u_{+}(D)+2i\epsilon}~,
\label{34}\\
&&\tilde{r}_L(\epsilon,D)=\frac{u_{-}(D)+2i\epsilon}
{u_{+}(D)+2i\epsilon}~,
\label{35}\\
&&\tilde{r}_R(\epsilon,D)=\frac{-u_{-}(D)+2i\epsilon}
{u_{+}(D)+2i\epsilon}~,
\label{36}
\eea
where 
\begin{equation}
u_{\pm}(D)=\Gamma_{\pm}(\Delta)({D/\Delta})^\alpha,
\label{37}
\end{equation}
and $\Gamma_\pm (\Delta)=\Gamma_1(\Delta)\pm\Gamma_2(\Delta)$ should
be found by matching onto Eqs.~(\ref{29b})--(\ref{29e}). The width of
a resonant tunneling peak $\Gamma(D)$ is thus given by $u_+(D)$.

\begin{figure}[ht]
\centerline{
\includegraphics[width=6.5cm]{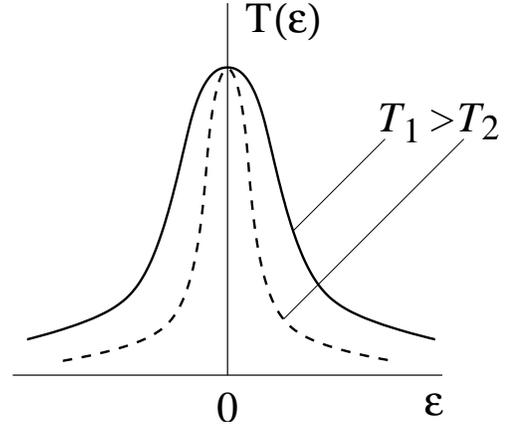}} 
\caption{\label{brw} Transmission coefficient ${\rm T}(\epsilon)$ for 
  temperatures $D_s\ll T\ll \Delta$. Two curves correspond to two
  temperatures: $T_1$ (solid) larger than $T_2$ (dashed). While the peak
  height is not renormalized by interactions, the resonance gets narrower with
  decreasing $T$, Eq.~(\ref{37a}).  }
\end{figure}

Note that the only condition we have assumed in the above derivation
is $D\gg \Gamma (D)$ with $D=\max\{|\epsilon|,{T}\}$, otherwise
$\epsilon$ in Eqs.~(\ref{34})--(\ref{36}) may be arbitrary. Thus,
Eqs.~(\ref{34})--(\ref{36}) give the shape of the $\epsilon$
dependence of fully renormalized amplitudes for the case of
temperature ${T}\gg \Gamma ({T})$ (with $T$ substituted
for $D$). In particular, Eq.~(\ref{37}) says that the width of the
resonance behaves as ${T}^\alpha$:
\be
\Gamma({T})=\Gamma_+(\Delta)({T}/\Delta)^\alpha.
\label{37a}
\ee 
As follows from Eq.~(\ref{34}), while the resonance becomes sharper with
decreasing $T$, the peak value of the transmission amplitude is not
renormalized (see Fig.~\ref{brw}), since the $D$ dependent factors cancel in
Eq.~(\ref{34}) at $\epsilon=0$. The absence of renormalization stems from the
vanishing of the sum $\tilde{r}_L(\epsilon,D)+\tilde{r}_R(\epsilon,D)$ in
Eq.~(\ref{32}) at $\epsilon=0$, which can be seen from
Eqs.~(\ref{35}),(\ref{36}).

We recognize Eqs.~(\ref{34})--(\ref{36}) as Breit-Wigner solutions that take
into account renormalization at $D\alt\Delta$. On the other hand, we have
already obtained Breit-Wigner formulas in Sec.~\ref{IIIc}, where the
renormalization has been carried out for $D\gg\Delta$, for $\epsilon$ close to
a resonance energy. In particular, the results of Sec.~\ref{IIIc} apply for
$|\epsilon|\ll\Delta$ if $\epsilon_0=0$. The matching of the two solutions at
$D\sim \Delta$ implies that Eqs.~(\ref{30}),(\ref{31}) are in fact valid in a
broader range of $D$, namely for $D\ll D_{r_{\rm min}}$, provided only that
one averages $r_R(\epsilon',D)$ over $\epsilon'$ together with the phase
factor $\chi_{\epsilon'}$. It follows that in the case of strong barriers
close to resonances the RG equations can be cast in the form
(\ref{32}),(\ref{33}) containing $t(\epsilon,D)$ and $r_{L,R}(\epsilon,D)$
only. Note also that for $D_0\alt \Delta$ the boundary conditions to
Eqs.~(\ref{32}),(\ref{33}) are fixed at $D=D_0$, which leads to the change
$\Delta\to D_0$ in Eq.~(\ref{37}).

At this point, one might be concerned about a possible contribution to
$t(\epsilon=0,D)$ from other resonances. Indeed, in the derivation of
Eq.~(\ref{34}), which gives no renormalization of $t(\epsilon=0,D)$,
we approximated $|r(\epsilon',D)|$ by unity for large $|\epsilon'|,D\gg
\Gamma(D)$. Corrections coming from other resonances are clearly small
in the parameter $\Gamma_{1,2}(D)/\Delta$ but one should check if they
might contribute to the renormalization of $t(\epsilon=0,D)$. Relaxing
the above approximation by allowing for resonant ``percolation" of
electrons through the barriers at $D\agt\Delta$ does give a
perturbative correction to the RG (\ref{32}):
\be
{\partial [\delta\tilde{t}(\epsilon,D)]\over \partial {\cal
L}_D}=-\alpha \tilde{t}(\epsilon,D){\pi u_-(D)\over
4\Delta}\,[\,\tilde{r}_L(\epsilon,D)-\tilde{r}_R(\epsilon,D)\,]~,
\label{38}
\ee
which, in contrast to Eq.~(\ref{32}), does not vanish at $\epsilon=0$
(unless the double barrier is symmetric: the correction is then always
zero). However, Eq.~(\ref{38}) tells us that the correction is
irrelevant since $u_-(D)$ itself scales to zero as $D^\alpha$. We thus
conclude that the ``single-peak approximation" of
Eqs.~(\ref{32}),(\ref{33}) correctly describes the renormalization of
the resonant amplitudes for all $D\gg\Gamma (D)$.

\subsection{Inside a peak: $D\ll D_s$}
\label{IIIe}

Now that we have integrated out all $D\gg u_+(D)$, let us continue
with the renormalization for $D$ inside a resonant tunneling peak. 
The point at which $D$ and $u_+(D)$ become equal to each other yields
a new characteristic scale $D_s$:
\be
D_s=\Gamma_+(\Delta)(D_s/\Delta)^\alpha=\Gamma_+(\Delta)
[\,\Gamma_+(\Delta)/\Delta\,]^{\alpha'},
\label{39}
\ee
where $\Gamma_+(\Delta)$ is obtained from
Eqs.~(\ref{29b})--(\ref{29e}) depending on the ratio of $D_{r_{1,2}}$
and $D_0$. To leading order in $\alpha$ the exponent
$\alpha'=\alpha/(1-\alpha)\to\alpha$. As will be seen below, the
significance of $D_s$ is that the width of the tunneling resonance
saturates with decreasing $D$ on the scale of $D_s$.

For $D\ll D_s$, the RG equations (\ref{30}),(\ref{31}) can be
simplified since the scattering amplitudes now depend on a single
variable, which is $D=\max \{|\epsilon|,{T}\}$. The averaged
reflection amplitudes $\overline{r_{L,R}}(D)$ coincide then with
$r_{L,R}(D)$, and the RG equations can be written in precisely the
same form as for a single impurity:
\be
\frac{\partial t(D)}{\partial {\cal L}_D} =
-\alpha t(D){\rm R}(D)~, \quad
\frac{\partial r_{L,R}(D)}{\partial {\cal L}_D} =
\ \alpha r_{L,R}(D){\rm T}(D)~,
\label{40}
\ee
with matching conditions at ${\cal L}_D=\ln (D_0/D_s)$. The difference
between the single structureless impurity and the resonance peak is
that in the latter case the ultraviolet cutoff is $D_s$. The fact that
the scattering amplitudes inside the peak are described by
Eqs.~(\ref{40}) was used also in Ref.~\onlinecite{nazarov02}.

In the preceding Secs.~\ref{IIIc},\ref{IIId}, asymmetry of the double barrier,
i.e., a possible difference between $t_1$ and $t_2$ was seen to determine the
amplitude and the width of a resonance peak but otherwise did not lead to any
qualitatively different consequences, as compared to the symmetric case.
However, for small $D\ll D_s$ the renormalization of the scattering amplitudes
is essentially different depending on whether the barriers are identical to
each other or not.

\begin{figure}[ht]
\centerline{
\includegraphics[width=8cm]{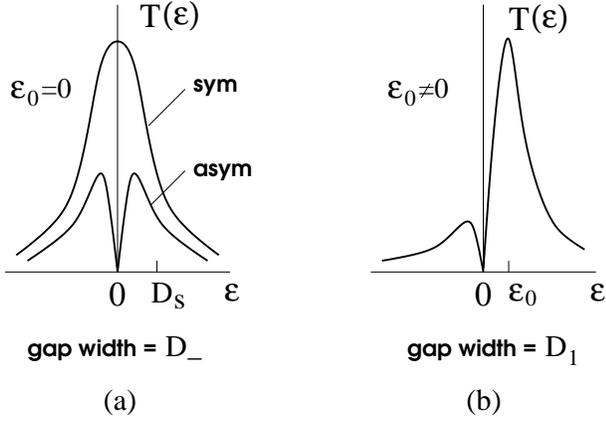}} 
\caption{\label{gap} Schematic summary of transmission 
  peak structure for zero temperature and strong barriers: (a) $\epsilon_0=0$,
  a single peak of ${\rm T}(\epsilon)$ for symmetric barriers with a width
  $D_s$ transforms into a double peak with a reduced height and a gap width
  $D_-< D_s$ for asymmetric barriers, Eqs.~(\ref{43}),(\ref{44}); (b)
  $0<|\epsilon_0|<D_s$, a single peak for symmetric barriers, centered at
  $\epsilon=\epsilon_0$, shows a dip at the Fermi energy, $\epsilon=0$, with a
  gap width $D_1< |\epsilon_0|$, Eq.~(\ref{46}). For $\epsilon_0\neq 0$ and
  asymmetric barriers, (a) describes the case of sufficiently small
  $\epsilon_0$, namely $D_-\gg D_1$, whereas (b) with a properly rescaled
  height describes the opposite case. The only effect of finite $T\alt D_s$ is
  to fill the gaps at $|\epsilon|\alt T$.}
\end{figure}

Consider first the symmetric case. Assuming, as in Sec.~\ref{IIId},
that the resonance energy $\epsilon_0=0$, we get the Breit-Wigner
formula with a width given by Eq.~(\ref{37}):
\be
\tilde{t}(D)=\frac{u_{+}(D)}{u_{+}(D)+2i\epsilon}~,
\label{41}
\ee
which turns out to be valid down to $D=0$. A remarkable consequence of
Eq.~(\ref{41}) is that the resonance in the symmetric case is perfect,
${\rm T}=1$ at $\epsilon=0$. While this is trivial for noninteracting
electrons, weak interaction in the Luttinger liquid is seen to
preserve the perfect transmission, in agreement with the result
obtained by a bosonic RG.\cite{kane92} So long as inelastic scattering
is not taken into account, the perfect transmission at $\epsilon=0$ is
not affected by finite temperature $T$, either, as can be seen
from Eq.~(\ref{41}) if one puts $D={T}$. However, the width of the
resonance does depend on $T$. At ${T}=0$, the width is finite
and given by $D_s$ (see Fig.~\ref{gap}a), which follows from Eq.~(\ref{41}) for
$D=|\epsilon|$. For ${T}\gg D_s$, the width obeys Eq.~(\ref{37a}).

The shape of the perfect resonance depends on the parameter
$T/D_s$. If ${T}\ll D_s$, the reflection probability as a function of
$|\epsilon|$ behaves near $\epsilon=0$ first as $\epsilon^2$ for
$|\epsilon|\ll {T}$ and then as $|\epsilon|^{2(1-\alpha)}$ for ${T}\ll
|\epsilon|\ll D_s$. For larger energies, the transmission probability
falls off with increasing $|\epsilon|$ as
\be
{\rm T}(\epsilon)=
(D_s/2|\epsilon|)^{2(1-\alpha)}~.
\label{42}
\ee 
This lineshape should be contrasted with the Lorentzian which
describes the transmission peak for ${T}\gg D_s$ up to
$|\epsilon|\sim {T}$, at which point a crossover to Eq.~(\ref{42})
occurs.

Let us now turn to the asymmetric case. Inspecting Eqs.~(\ref{40}), we
see that a new characteristic scale $D_-$
emerges:
\be
D_{-}= D_s\left(\frac{|\Gamma_- (\Delta)|}{\Gamma_+ 
(\Delta)}\right)^{1/\alpha},
\label{43}
\ee
which coincides with $D_s$ for strongly asymmetric barriers but
vanishes for symmetric ones. For ${T}\agt D_-$ we get the same
results for ${\rm T}(\epsilon)$ as in the symmetric case, only with an
overall factor of 
\be
\lambda ={\Gamma_{+}^2(\Delta) -\Gamma_{-}^2(\Delta)\over
\Gamma_+^2(\Delta)}={4\Gamma_1(\Delta)\Gamma_2(\Delta)\over [\,
\Gamma_1(\Delta)+\Gamma_2(\Delta)\,]^2}~.
\label{43a}
\ee
However, for ${T}\ll
D_-$ a new feature in the behavior of ${\rm T}(\epsilon)$ shows up, namely a
power-law falloff with decreasing $|\epsilon|$. The function ${\rm T}(D)$,
as obtained from Eqs.~(\ref{40}), for $D\ll D_s$ reads
\be
{\rm T}(D)={\lambda (D/D_s)^{2\alpha}\over
1-\lambda\,[\,1-(D/D_s)^{2\alpha}\,]}~.
\label{44}
\ee
One sees that ${\rm T}(\epsilon)$ behaves as (Fig.~\ref{gap}a)
\be
{\rm T}(\epsilon)=\lambda(|\epsilon|/D_-)^{2\alpha}
\label{45}
\ee 
in the interval ${T}\ll |\epsilon|\ll D_-$ and saturates at smaller
energies: ${\rm T}(\epsilon=0)=\lambda({T}/D_-)^{2\alpha}$. Thus, in the limit
${T}\ll D_-$, the resonant transmission probability as a function of
$\epsilon$ exhibits a double-peak structure, see Fig.~\ref{gap}a. If, however,
the barriers are only slightly asymmetric, the gap near $\epsilon=0$ develops
in a range of $\epsilon$ which is much narrower than the width of the
resonance peak. Specifically, ${\rm T}(\epsilon)$ first grows up wit
increasing $|\epsilon|$ for ${T}\ll |\epsilon|\ll D_-$, then there is a
plateau with an energy independent transmission for $D_-\ll |\epsilon|\ll
D_s$, and ${\rm T}(\epsilon)$ starts to fall off as $|\epsilon|$ is further
increased.

In the above, we analyzed the behavior of ${\rm T}(\epsilon)$ for the
resonance energy $\epsilon_0=0$, i.e., when it coincides with the
Fermi level. Let us now examine the case $\epsilon_0\neq 0$. Again,
let the barriers first be symmetric. Then, in Eq.~(\ref{41}), the
resonant denominator changes to $u_+(D)+2i(\epsilon-\epsilon_0)$ and
$D$ is, as before, $\max\{|\epsilon|,{T}\}$. The innocent
looking shift $\epsilon\to\epsilon-\epsilon_0$ leads at ${T}=0$ to
dramatic consequences for transmission at the Fermi energy,
$\epsilon=0$. Namely, ${\rm T}(\epsilon)$ is now seen to vanish at
$\epsilon=0$ and zero $T$, whatever $\epsilon_0$ unless it is
exactly zero. A new characteristic scale $D_1$ becomes relevant at
$\epsilon_0\neq 0$: it is defined by $u_+(D_1)=2|\epsilon_0|$, which
is rewritten as
\be
D_1=D_s(2|\epsilon_0|/D_s)^{1/\alpha}~.
\label{46}
\ee
The significance of the energy $D_1$ is that the width of the gap in
the dependence of ${\rm T}(\epsilon)$ around $\epsilon=0$ at ${T}=0$ is
given by $D_1$ for $|\epsilon_0|\alt D_s$.  Note that $D_1\ll
|\epsilon_0|$ for $|\epsilon_0|\ll D_s$.

The shape of the resonant peak as a function of $\epsilon$ changes in
an essential way for $|\epsilon_0|\alt D_s$. Specifically, if
${T}\gg D_1$, the changes are weak; however, for ${T}\ll D_1$
a range of $\epsilon$ arises, ${T}\ll |\epsilon|\ll D_1$, within
which ${\rm T}(\epsilon)$ behaves as (Fig.~\ref{gap}b)
\be
{\rm T}(\epsilon)=(|\epsilon|/D_1)^{2\alpha}.
\label{47}
\ee
The power-law falloff (\ref{47}) saturates close to the Fermi level at
${\rm T}(\epsilon=0)=({T}/D_1)^{2\alpha}$. 

We thus see that the width of the resonance in the transmission
through a symmetric barrier exactly at the Fermi energy ${\rm
T}(\epsilon=0)$ as a function of $\epsilon_0$ vanishes as ${T}\to
0$. On the other hand, the width of the resonance in the transmission
at $\epsilon_0=0$ as a function of $\epsilon$ is finite even at $T=0$
and is given by $D_s$. This peculiar feature is in sharp contrast to
the resonant tunneling of noninteracting electrons, for which the two
widths are the same.

For asymmetric barriers, ${\rm T}(\epsilon)$ does not change substantially
with increasing $|\epsilon_0|$ as long as $D_1\ll D_-$ and is given by the
formulas for symmetric barriers with an overall reduction of the transmission
probability by a factor of $\lambda$ (\ref{43a}) otherwise (see
Fig.~\ref{gap}).

\subsection{Weak barriers}
\label{IIIf}

In Secs.~\ref{IIIc}--\ref{IIIe}, we have assumed that $D_{r_{\rm
min}}\gg\Delta$, which means that even if the impurities are initially
(at $D=D_0$) weak, they get strong in the process of renormalization
before $D$ becomes equal to $\Delta$. Under this assumption, we have
sharp resonant peaks close to the Fermi level and the bare scattering
amplitudes only rescale [according to Eqs.~(\ref{29b})--(\ref{29e})]
parameters in otherwise general formulas for the resonant tunneling.
Let us now examine the resonant transmission in the case of at least
one barrier being initially so weak that the renormalization does not
make it strong at $D\sim\Delta$.

We begin with the case of both barriers characterized by
$D_{r_{1,2}}\ll \Delta$. The total reflection coefficient
$r_L(\epsilon,D)$ as obtained from Eqs.~(\ref{21}),(\ref{22}) in the
limit $|r_L|\ll 1$ is given by
\be
r_L(\epsilon,D)=(r_1-r_2\chi_\epsilon)(D_0/D)^\alpha
\label{48}
\ee
with $D=\max\{|\epsilon|,{T}\}$.
It is worth noting that, due to the oscillating factor $\chi_\epsilon$
(\ref{17a}), it is not possible to derive Eq.~({\ref{48}) from
Eqs.~(\ref{19}),(\ref{20}) even in the simplest case of a weak double
barrier.

Suppose first that the barrier is symmetric and
$\epsilon_0=0$. Then Eq.~(\ref{48}) simplifies  to
\be
{\rm R}(\epsilon,D)=2[\,1-\cos(2\pi\epsilon/\Delta)\,]\,
(D_r/D)^{2\alpha}.
\label{49}
\ee One sees that reflection is enhanced by interaction, but the reflection
coefficient is always small, ${\rm R}\ll 1$ for any $\epsilon$, provided
$D_r\ll\Delta$. No sharp features in the $\epsilon$ dependence of the
scattering amplitudes emerge around the Fermi energy (see Fig.~\ref{weak}a).

\begin{figure}[ht]
\centerline{
\includegraphics[width=8cm]{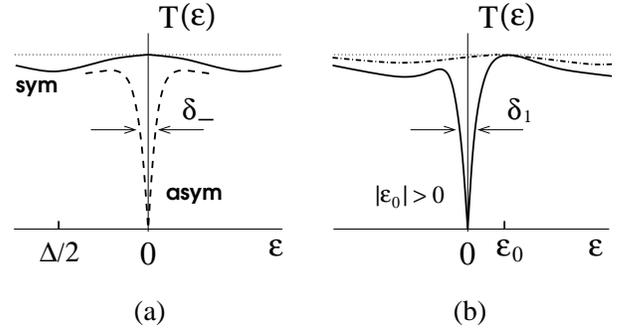}} 
\caption{\label{weak}  Schematic behavior of the transmission coefficient
  for zero temperature and weak barriers, Eq.~(\ref{55}): (a) for
  $\epsilon_0=0$, the modulation of ${\rm T}(\epsilon)$ for both symmetric
  (solid) and asymmetric (dashed) barriers is strongly increased. However, for
  symmetric barriers ${\rm T}(\epsilon)$ remains close to unity for all
  $\epsilon$, while there appears a dip of width $\delta_-<D_r$,
  Eqs.~(\ref{51}),(\ref{52}), around $\epsilon=0$ for asymmetric barriers; (b)
  for $\epsilon_0\neq 0$ and symmetric barriers, the bare transmission
  (dash-dotted) is strongly renormalized (solid) due to interactions and
  exhibits a gap of width $\delta_1< D_r$, Eq.~(\ref{54}), around the Fermi
  level.  For $\epsilon_0\neq 0$ and asymmetric barriers, (a) describes the
  case of sufficiently small $\epsilon_0$, namely $\delta_-\gg \delta_1$,
  whereas (b) with a properly rescaled height describes the opposite case.
  Similarly to Fig.~\ref{gap}, finite $T$ fills the gaps at $|\epsilon|\alt
  T$.}
\end{figure}

It is now instructive to introduce a weak asymmetry ${\rm R}_- =|{\rm R}_2-{\rm
  R}_1|$, such that ${\rm R}_-\ll {\rm R}\simeq {\rm R}_{1,2}$. Given that the
asymmetry is weak, it can manifests itself only at small energies. Expanding
Eq.~(\ref{48}) about $\epsilon=0$, we get for $|\epsilon|\ll \Delta$:
\be
{\rm R}(\epsilon,D)=\left[\left({{\rm R}_-\over 2{\rm R}}\right)^2+
\left({2\pi\epsilon\over \Delta}\right)^2\right]
\left({D_r\over D}\right)^{2\alpha}.
\label{50}
\ee 
As can be seen from Eq.~(\ref{50}), asymmetry sets two new characteristic
scales of energy: $({\rm R}_-/{\rm R})\Delta$ and a smaller scale
\be
\delta_-=D_r({\rm R}_-/2{\rm R})^{1/\alpha}.
\label{51}
\ee
Provided that temperature ${T}\ll ({\rm R}_-/{\rm R})\Delta$, the
reflection coefficient starts to grow with approaching the Fermi level
at $|\epsilon|\sim ({\rm R}_-/{\rm R})\Delta$. The enhancement of
reflection is cut off by temperature before $R$ becomes of order unity
if $T$ is not too low, specifically if $\delta_-\ll {T}$. However, if
$T\ll \delta_-$, then reflection gets strong at $|\epsilon|\sim
\delta_-$. To describe the scattering probabilities at $|\epsilon|\alt
\delta_-$, one should solve Eqs.~(\ref{40}), derived in the same way
it was done in Sec.~\ref{IIIe}, now with matching onto the
perturbative (in ${\rm R}_{1,2}$) solution (\ref{48}) anywhere in the
region $\delta_-\ll |\epsilon|\ll \Delta$. For $D\ll ({\rm R}_-/{\rm
R})\Delta$ the solution reads:
\be
{\rm T}(D)=1/\,[\,1+(\delta_-/D)^{2\alpha}]~.
\label{52}
\ee
Thus, the weak double barrier remains slightly reflecting after the
renormalization provided that ${T}\gg \delta_-$. However, for both
${T},|\epsilon|\ll \delta_-$  the
transmission probability is small: within the range ${T}\ll
|\epsilon|\ll \delta_-$, ${\rm T}(\epsilon)$ behaves as (Fig.~\ref{weak}a)
\be
{\rm T}(\epsilon)=(|\epsilon|/\delta_-)^{2\alpha}~,
\label{53}
\ee
and saturates for smaller $|\epsilon|$ at
${\rm T}(\epsilon=0)=(T/\delta_-)^{2\alpha}$. Comparing
Eqs.~(\ref{53}),(\ref{45}) with each other, we see that the energy
$\delta_-$ is a counterpart of $D_-$ for the case of a weak barrier.

Generalizing to $\epsilon_0\neq 0$, we have for
$|\epsilon_0|\ll\Delta$ a shift $\epsilon\to\epsilon-\epsilon_0$ in
Eq.~(\ref{50}), while $D=\max\{|\epsilon|,{T}\}$.
A new energy scale $\delta_1\ll \epsilon_0$ appears, at which 
the reflection coefficient becomes of order unity in the
symmetric case:
\be
\delta_1=D_r\,(2\pi|\epsilon_0|/\Delta)^{1/\alpha},
\label{54}
\ee 
analogous to $D_1$ in Eq.~(\ref{46}) for tunneling barriers. The energy
$\delta_1$ gives the width of the gap in the transmission probability at the
Fermi level at ${T}=0$. For ${T}\ll |\epsilon|\ll \delta_1$, we get a
power-law vanishing of ${\rm T}(\epsilon)=(|\epsilon|/\delta_1)^{2\alpha}$
with decreasing $|\epsilon|$ (see Fig.~\ref{weak}b) and a saturation for
smaller $|\epsilon|$ at $(T/\delta_1)^{2\alpha}$. A general expression for the
scattering probabilities, valid for arbitrary $D_{r_{1,2}}\ll \Delta$ and
$|\epsilon|,D\ll\Delta$, can be obtained from Eqs.~(\ref{40}): 
\bea {{\rm
    R}(\epsilon,D)\over {\rm T}(\epsilon,D)}&=&\left[\left({\rm R}_1^{1/2}
    -{\rm R}_2^{1/2}\right)^2
\right. \label{55} \\
&+&\left. ({\rm R}_1{\rm R}_2)^{1/2}\left({2\pi\over\Delta}\right)^2
  (\epsilon-\epsilon_0)^2 \right]\left({D_0\over D}\right)^{2\alpha}.
\nonumber 
\eea 
Equation (\ref{55}) reproduces Eqs.~(\ref{50})--(\ref{54}) in
the corresponding limits.

We conclude that if the barriers are symmetric but $\epsilon_0$ is nonzero, or
if the barriers are asymmetric, the transmission probability vanishes
(Fig.~\ref{weak}) at the Fermi level in the limit ${T}\to 0$. We will see in
Sec.~\ref{IV} that these features lead to the emergence of a sharp peak in the
low-temperature conductance as a function of $\epsilon_0$ even for two weak
impurities, provided only that they are slightly asymmetric.

Finally, when two strongly asymmetric barriers are located nearby, so
that at $D\sim\Delta$ one barrier is strongly reflecting whereas the
other is still weak, the effect of the latter on the transmission
probability remains small for any $D$. Let us take the example
$D_{r_1}\gg \Delta$ and $D_{r_2}\ll \Delta$. Then we get for
$D_{r_1}\gg D\gg\Delta$
\be
{\rm T}(\epsilon,D)
=(D/D_{r_1})^{2\alpha}\,[\,1+2(D_{r_2}/D)^\alpha\cos\theta\,]~, 
\label{56}
\ee
where $\theta=2\pi(\epsilon-\epsilon_0)/\Delta$. One sees that the
presence of the weak impurity only leads to a weak modulation with
changing $\epsilon$. For $D\alt \Delta$, the independent
renormalization of the weaker impurity is suppressed by reflection
from the strong barrier and ${\rm T}(D)$ behaves as $(D/D_{r_1})^{2\alpha}$
down to $D=0$.

\section{Conductance peak}
\label{IV} 

The solution to the problem of transmission through a double barrier
given in the preceding sections allows us to examine the linear
conductance of the system $G(\epsilon_0,{T})$ as a function of
temperature $T$ and the energy distance between the Fermi level and a
resonance level $\epsilon_0$. Recall that we have studied the elastic
transmission of interacting electrons, i.e., the energy $\epsilon$ of
an incident electron before and after the transmission is the same,
while in the process of scattering off the barrier $\epsilon$ is not
conserved due to interaction with other electrons both in the
particle-hole and Cooper channels. At finite $T$, there are also
inelastic processes, characterized by the inelastic scattering length
$L_{\rm in}$. Neglecting the inelastic scattering is legitimate if
$L_{\rm in}\gg L_{T}\sim v_F/{T}$, which is satisfied in the present
problem for weak interaction $\alpha\ll 1$. On the other hand, it is
worthwhile to note that the very formulation of the scattering problem
in the interacting case even for elastic scattering is a delicate
issue if one keeps scaling exponents of higher order than linear in
$\alpha$.\cite{inprep} Also, interaction-induced current-vertex
corrections in the Kubo formula in the presence of impurities and
interaction are governed by exponents of higher order in
$\alpha$. Here, we avoid these complications by treating the
scattering problem within the one-loop approximation, i.e., keeping
only first order terms in the exponents. Under these
conditions, one can use the Landauer-B\"uttiker formalism
relating the conductance and the transmission probability. The
conductance $G(\epsilon_0,{T})$ in units of $e^2/h$ reads
\be
G(\epsilon_0,{T})= \int\! d\epsilon\,\, {\rm T}(\epsilon)\,(-\partial
n_F/\partial\epsilon)~. 
\label{57} 
\ee
We are interested in the low-temperature regime with ${T}\ll
\Delta$, otherwise we intend to keep $T$ arbitrary, i.e.,
$T$ may be as small as zero.

Below we analyze various limiting cases. As we have seen in Sec.~\ref{III},
the strength of interaction, the strength of the barriers, and the degree of
asymmetry set a number of characteristic energy scales which yield a variety
of different regimes in the temperature and energy dependence of the
transmission probability ${\rm T}(\epsilon,{T})$. By means of Eq.~(\ref{57}),
these regimes manifest themselves in the behavior of the conductance
$G(\epsilon_0,{T})$, which is a directly measurable quantity.

Before turning to the calculation of $G(\epsilon_0,{T})$, let us briefly
discuss the restrictions and possible implementations of our model. Even
though we are dealing with the case of weak interaction, $\alpha\ll 1$, the
product $\alpha\ln (D_0/\Delta)$ may be large in small quantum dots, so that
the renormalization of scattering amplitudes on scales $D\gg\Delta$ is
necessary, as has been done in Sec.~\ref{IIIc}. Experimentally, as discussed
in Sec.~\ref{I}, the parameter $\alpha$ is typically not small in carbon
nanotubes [in Refs.~\onlinecite{bockrath99,yao99}, the value of $\alpha$
extracted according to Eq.~(\ref{alphaend}) from the measured $\alpha_e\sim
0.6-1.0$ is $\alpha\sim 1.3-3.0$)].  On the other hand, in single-mode
semiconductor quantum wires the strength of interaction is typically smaller
(in Ref.~\onlinecite{auslaender00}, $\alpha_e\sim 0.2-0.5$ corresponds
[Eq.~(\ref{alphaend})] to $\alpha\sim 0.3-0.8$) and can be more easily made
small through screening by nearby metallic gates.  In weakly interacting
wires, our results would be applicable directly. At the same time, our theory
captures much of the essential physics of strongly [with $\alpha$ defined in
Eq.~(\ref{2}) of order unity] correlated wires too, as follows from the
comparison with the known results obtained by different
methods.\cite{kane92,furusaki93,furusaki98}

As for the impurity strength, both transport and tunneling experiments on
quantum wires with imperfections have been so far focused on the case of
initially strong inhomogeneities (impurities, artificially created tunneling
barriers, or non-adiabatic contacts). Sections \ref{IIIc}-\ref{IIIe} describe
this situation in detail.  Moreover, as shown in Secs.~\ref{IIIc},\ref{IIIf},
weak inhomogeneities, which are potentially realizable in a controllable way
both in semiconductor quantum wires and carbon nanotubes, are of special
interest, since the Coulomb interaction transforms two weak impurities into a
quantum dot with a pronounced resonance structure. Finally, a relatively
strong asymmetry of the quantum dot appears to be inevitably present in the
experimental setups of Refs.~\onlinecite{auslaender00,postma01}.

\subsection{Strong barriers}

Consider first the case of strong barriers (more precisely, the bare
transmission through the barriers may be high, ${\rm T}_{1,2}\simeq 1$, but we
assume that the barriers get strong before the RG flow parameter $D$ equals
the single-particle energy spacing inside the dot, $\Delta$), i.e., $D_{r_{\rm
    min}}=D_0(\min\{{\rm R}_1,{\rm R}_2\})^{1/2\alpha}\gg \Delta$. Then we
have a sharp peak of the transmission probability centered at
$\epsilon=\epsilon_0$ whose width is $\max\{D_s,\Gamma({T})\}\ll \Delta$,
where $D_s$ and $\Gamma ({\rm T})$ are defined in Eqs.~(\ref{37a}),(\ref{39}).
In other words, the width of the peak in ${\rm T}(\epsilon,{T})$ is
$\Gamma({T})=D_s(T/D_s)^\alpha$ for ${T}\gg D_s$, whereas for smaller ${T}\ll
D_s$ the width is of order $D_s$ and does not depend on $T$.

\subsubsection{${T}\gg D_s$, Sequential tunneling}

For $T\gg D_s$, the shape of the
conductance peak is given by (see Fig.~\ref{cond})
\be
G(\epsilon_0,{T})={\zeta \,G_p \over \cosh^2(\epsilon_0/2{T})}~,
\label{58}
\ee
where the peak value of the conductance
\be
G_p=\pi\lambda \Gamma({T})/8{T}~,
\label{59}
\ee with $\lambda$ defined in Eq.~(\ref{43a}), and
$\zeta=(\max\{|\epsilon_0|,T\}/T)^\alpha$.  The width of the conductance peak
$w$ is of order $T$, as for noninteracting electrons; however, the power-law
behavior of $G_p({T})$ is seen to be modified by interaction, in accordance
with the results derived in Refs.~\onlinecite{furusaki93,furusaki98}.  Note
that the scaling of $G_p\propto {T}^{\alpha-1}$ is governed by the
single-particle density of states $\rho_e({T})$ for tunneling into the end of a
Luttinger liquid, namely $G_p\propto \rho_e({T})/{T}$.\cite{fisher96} 

We recognize Eqs.~(\ref{58}),(\ref{59}) as the conventional sequential
tunneling formulas, but with a $T$ dependent resonance width $\Gamma ({T})$.
Far in the wings of the resonance the exponential falloff (\ref{58}),
$G(\epsilon_0,T)\sim \lambda T^{-1}\Gamma(|\epsilon_0|)\exp
(-|\epsilon_0|/T)$, crosses over onto the cotunneling (determined by
the processes of fourth order in tunneling amplitudes) power law
$G(\epsilon_0,{T})=\lambda \Gamma^2(T)/4\epsilon^2_0$, as usual. The crossover
between the sequential tunneling and cotunneling regimes occurs at
$|\epsilon_0|\simeq T\ln\,[\,T/\Gamma (T)\,]$. In fact, this formula is
valid\cite{furusaki98} for an arbitrary strength of interaction with $\Gamma
(T)\propto \rho_e (T)\propto T^{\alpha_e}$, where $\alpha_e$ (equal to
$\alpha$ for a weak interaction) is the end-tunneling exponent
(\ref{alphaend}). It is worthwhile to note that for strong enough interaction
(namely for $\alpha_e>1$) the sequential mechanism of tunneling is effective
for the resonance peak for all $T$, down to $T=0$.\cite{furusaki98} Moreover,
for $\alpha_e>1$ the crossover to the cotunneling regime shifts towards larger
$|\epsilon_0|$ with increasing strength of interaction.

As demonstrated in Ref.~\onlinecite{furusaki98},
Eqs.~(\ref{58}),(\ref{59}) can be obtained from a classical kinetic
equation\cite{beenakker91} for occupation numbers characterizing the
state of the quantum dot.  We get the same results from the fermionic
RG equations.  It is worth mentioning that, although
Eqs.~(\ref{58}),(\ref{59}) follow from a classical kinetic
equation\cite{beenakker91} which involves diagonal elements of the
density matrix only, the transmission in our derivation is fully
coherent. The fact that in the high-temperature limit it can also be
described in terms of the classical kinetic equation merely means that
quantum corrections to the kinetic equation may be neglected for small
$\Gamma({T})/{T}$ [whereas the quantum suppression of the tunneling
density of states $\rho_e(T)$ may be treated in this equation in a
phenomenological way].

\begin{figure}[ht]
\includegraphics[width=4cm]{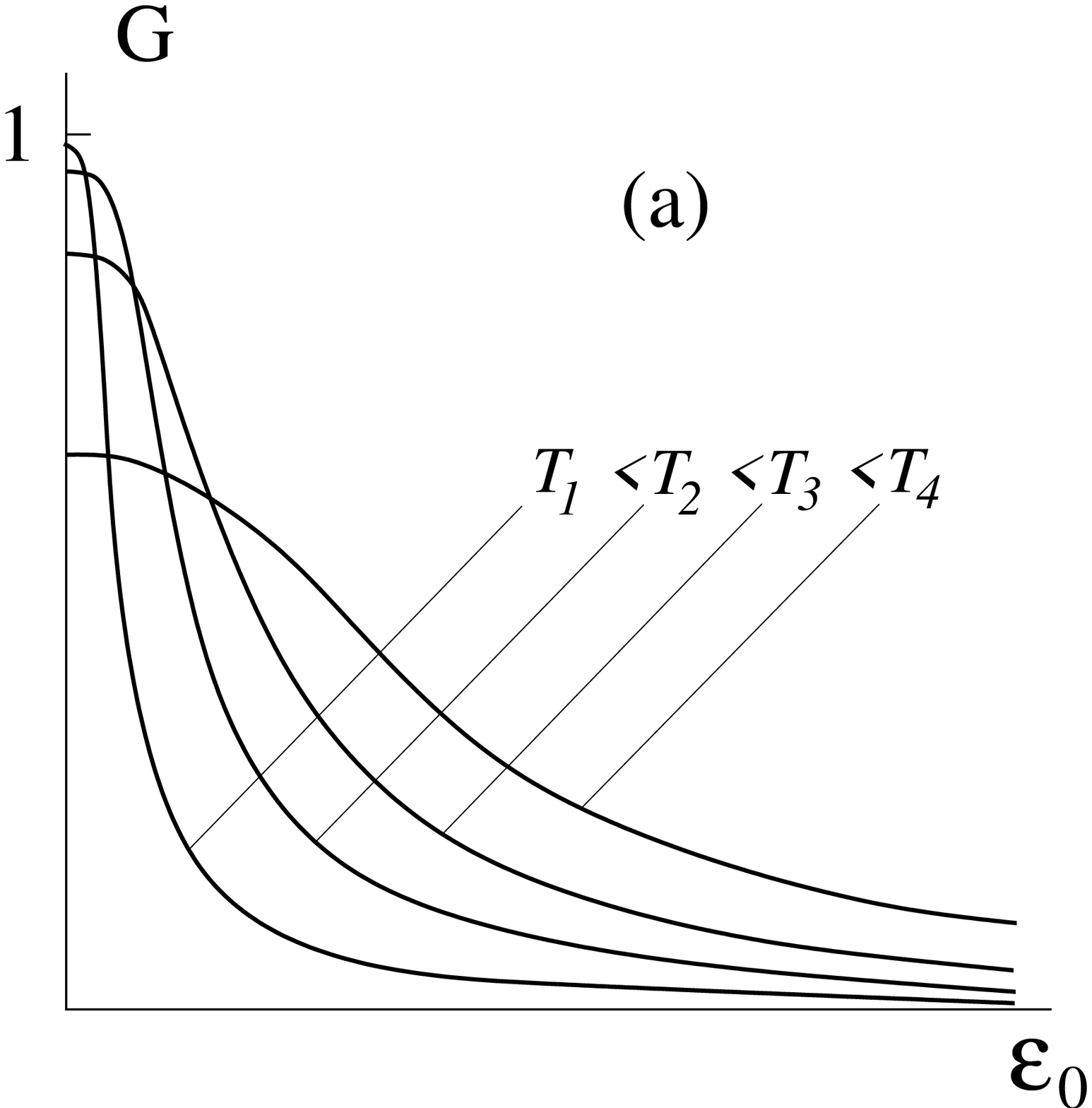}
\includegraphics[width=4cm]{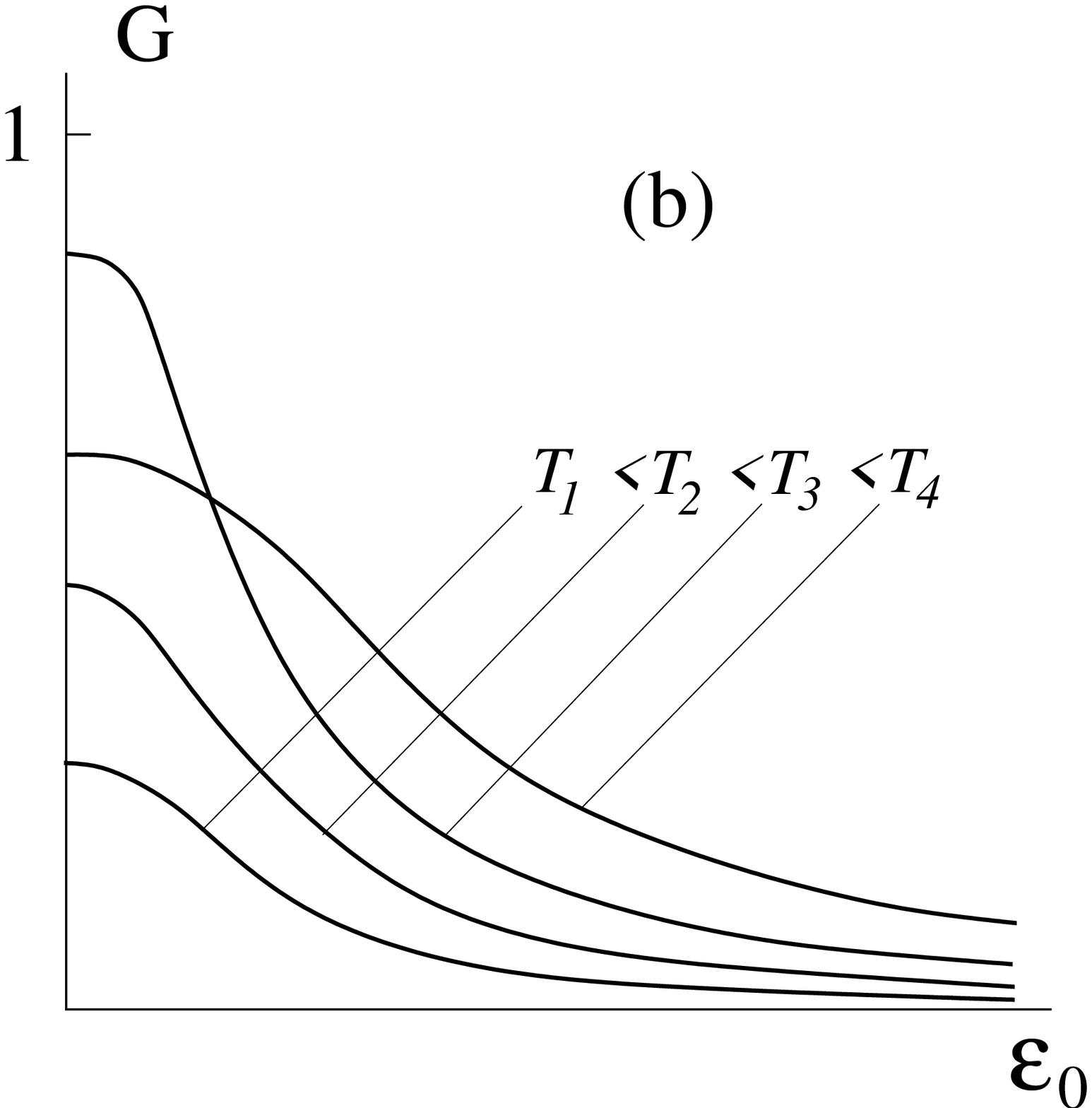}
\caption{\label{cond} 
  Strong barriers: Qualitative dependence of the conductance $G$ (in units of
  $e^2/h$) on $\epsilon_0$, the energy difference between the Fermi energy and
  the resonance level, for different temperatures $T$ and weak interaction.
  (a) Resonance contour for {\it symmetric} barriers. The peak conductance
  $G_p$ grows [Eq.~(\ref{59})] with decreasing $T$ and saturates
  [Eq.~(\ref{60})] at $G_p=1$ as $T\to 0$.  The width of the resonance shrinks
  to zero, according to Eqs.~(\ref{58}),(\ref{61}). (b) Resonance contour for
  {\it asymmetric} barriers. The dependence of $G_p$ on $T$ is non-monotonic:
  $G_p$ grows [Eq.~(\ref{59}), curves $T_4$ and $T_3$] with decreasing $T$ for
  large $T$ and goes down [Eq.~(\ref{63}), curves $T_2$ and $T_1$] for
  $T<D_-$. The width of the resonance decreases [Eq.~(\ref{58})] with lowering
  $T$ and saturates [Eq.~(\ref{63})] as $T\to 0$.}
\end{figure}

\subsubsection{${T}\ll D_s$, Symmetric barriers}

Let us now turn to low temperatures ${T}\ll D_s$, where processes of all
orders in the tunneling amplitudes are important. Consider first the
symmetric case (Fig.~\ref{cond}a). The main contribution to the integral over
$\epsilon$ in Eq.~(\ref{57}) comes from $|\epsilon|\sim {T}\ll D_s$, so that
the shape of the conductance peak is a Lorentzian: 
\be
G(\epsilon_0,{T})={\Gamma^2(T)\over \Gamma^2(T)+4\epsilon_0^2}~.
\label{60}
\ee
We see that the height of the peak $G_p=1$ and the width 
\be
w=D_s({T}/D_s)^\alpha
\label{61}
\ee 
exhibits a power-law temperature dependence with an exponent depending on
the strength of interaction. The vanishing of $w$ as ${T}\to 0$ should be
contrasted with the behavior of the peak in ${\rm T}(\epsilon,T)$
(Fig.~\ref{gap}), whose width is $D_s$ for low $T$. In the limit ${T}\to 0$,
the conductance peak becomes infinitely narrow but the resonance at
$\epsilon_0=0$ persists down to ${T}=0$, in accordance with
Ref.~\onlinecite{kane92}.  We thus confirm the persistence\cite{kane92} of the
perfect resonance at $T=0$ by means of the fermionic RG. For finite $T\ll
D_s$, there is a small correction $1-G_p\sim (T/D_s)^{2(1-\alpha)}$, which
comes from the non-perfect transmission for finite $\epsilon$ at
$\epsilon_0=0$ after the thermal averaging (\ref{57}).

\subsubsection{${T}\ll D_s$, Asymmetric barriers}
 
We recall that the renormalization in the asymmetric case is governed by the
scale $D_-$ [defined in Eq.~(\ref{43})] which describes the degree of
asymmetry. The double-peak structure of the transmission coefficient as a
function of $\epsilon_0$ for ${T}\ll D_-$ translates into a complete vanishing
of the conductance peak at $T\to 0$. Specifically, for ${T}\gg D_-$ the
conductance $G(\epsilon_0,{T})$ is given by Eq.~(\ref{60}) for symmetric
barriers with an overall factor of $\lambda$. However, for ${T}\ll D_-$ the
transmission at the Fermi level falls off with decreasing $T$ (Fig.~\ref{gap})
and so does the conductance peak (see Fig.~\ref{cond}b): 
\be
G(\epsilon_0,{T})={\lambda ({T}/D_s)^{2\alpha}\over
  (D_-/D_s)^{2\alpha}+(2\epsilon_0/D_s)^2}~,
\label{62}
\ee  
which gives
\be
G_p=\lambda \left({{T}\over D_-}\right)^{2\alpha}~,\quad
w=D_s{|\Gamma_-(\Delta)|\over\Gamma_+(\Delta)}~,
\label{63}
\ee
where $\Gamma_{\pm}(\Delta)$ are defined below Eq.~(\ref{37}).
Thus, at small ${T}\ll D_-$, the height of the conductance peak goes
down as $T$ decreases, whereas the width of the peak does not depend
on $T$ any longer. This kind of behavior of the resonance peak was
predicted in Ref.~\onlinecite{kane92}: $G_p({T})$ scales as
$\rho^2({T})$, as for a single impurity. The role of asymmetry was
emphasized and expressions similar to Eqs.~(\ref{63}) were obtained in
Ref.~\onlinecite{nazarov02}. However, the authors of
Ref.~\onlinecite{nazarov02} do not distinguish between the scales
$D_-$ and $D_s$ (the latter is denoted $\tilde\epsilon$ there).

\subsection{Weak barriers}

Consider now the case of weak barriers, i.e., $D_{r_{1,2}}\ll \Delta$. Naively
one could think that scattering on weak barriers cannot possibly yield a sharp
peak of conductance. Indeed, the transmission probability as a function of
$\epsilon$ (Fig.~\ref{weak}) does not have any peak at $\epsilon=\epsilon_0$,
in contrast to the case of resonant tunneling. At high $T\gg D_r$,
$G(\epsilon_0,T)$ is a weakly oscillating (with a period $\Delta$) function of
$\epsilon_0$. The only difference with the non-interacting case is an enhanced
amplitude of the oscillations.

Let us show that in fact the interaction-induced vanishing of ${\rm
  T}(\epsilon)$ at the Fermi energy $\epsilon=0$ for ${T}=0$ does lead to a
narrow Lorentzian peak of $G(\epsilon_0,{T})$ (see Fig.~\ref{weakcond}),
provided that $T$ is low enough and the barriers are not too asymmetric.

\subsubsection{Symmetric barriers}

Integration
(\ref{57}) of the transmission probability (\ref{55}) for symmetric
barriers yields
\be
G(\epsilon_0,{T})=\left[1+\left({2\pi\epsilon_0\over
\Delta}\right)^2\left({D_r\over {T}}\right)^{2\alpha}\right]^{-1}~,
\label{64}
\ee 
which indeed describes a Lorentzian peak (Fig.~\ref{weakcond}a) with the
height $G_p=1$ and the width
\be
w={\Delta\over\pi}\left({{T}\over D_r}\right)^\alpha.
\label{65}
\ee
It follows that the peak is narrow, $w\ll\Delta$, provided that $T\ll D_r$. In
the limit ${T}\to 0$, the width of the peak is infinitesimally small.
Similarly to the case of resonant tunneling, the resonance at $\epsilon_0=0$
remains perfect in the presence of interaction at $T=0$. At finite $T$, there
is a small correction 
\be
1-G_p\sim (T/\Delta)^2(D_r/T)^{2\alpha}~.
\label{65a}
\ee
Equation (\ref{65a}) describes also the high-$T$ behavior of $G_p$ in the case
of slightly asymmetric barriers (see below).

\begin{figure}[ht]
\includegraphics[width=4cm]{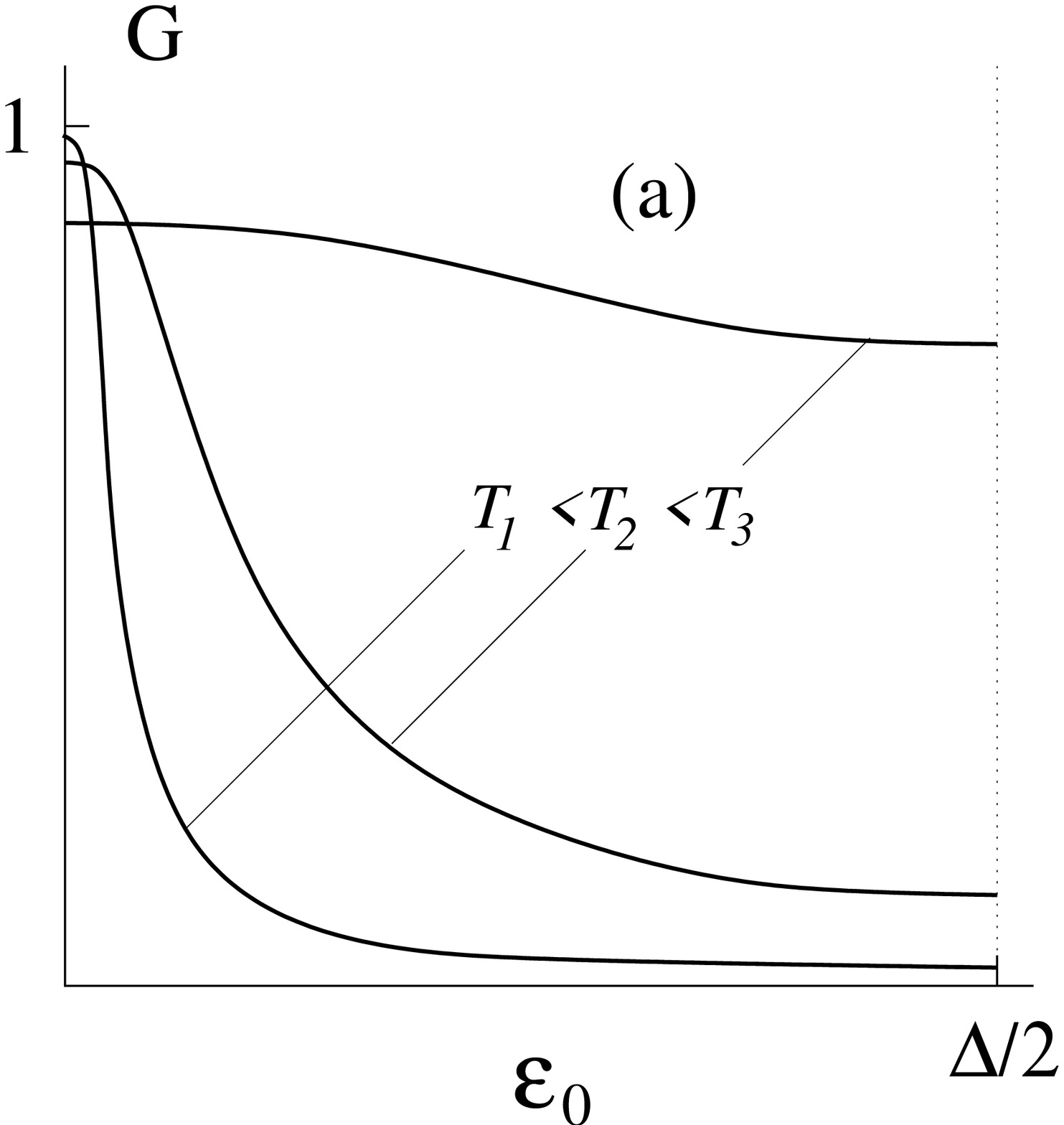}
\includegraphics[width=4cm]{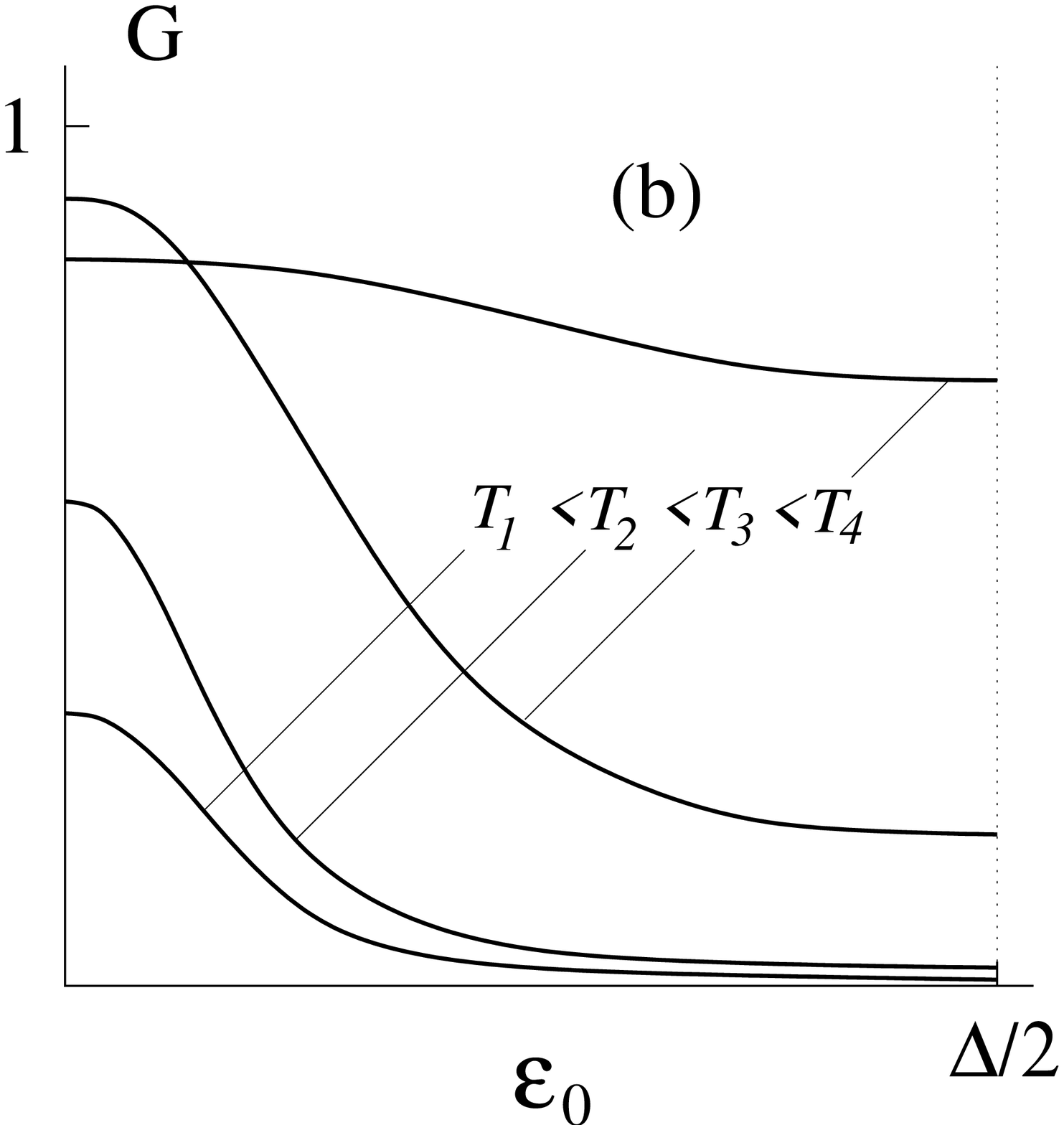}
\caption{\label{weakcond} 
  Weak barriers: Qualitative dependence of the conductance $G$ (in units of
  $e^2/h$) on the resonance energy $\epsilon_0$ for different temperatures $T$
  and weak interaction. (a) {\it Symmetric} barriers. The value of
  $G(\epsilon_0=0)$ grows [Eq.~(\ref{65a})] with decreasing $T$ and saturates
  [Eq.~(\ref{65})] at $G_p=1$ as $T\to 0$. The width of the resonance shrinks
  to zero, according to Eqs.~(\ref{64}),(\ref{65}). The resonance peak becomes
  visible at $T\ll D_r$ (curves $T_2$ and $T_1$). (b) {\it Asymmetric}
  barriers. The dependence of $G(\epsilon_0=0)$ on $T$ is non-monotonic: it
  grows [Eq.~(\ref{65a})] with decreasing $T$ for large $T$ and goes down
  [Eq.~(\ref{67})] for $T<\delta_-$. The width of the resonance decreases
  [Eq.~(\ref{65}), similarly to the symmetric case] with lowering $T$ and
  saturates [Eq.~(\ref{67}), curves $T_2$ and $T_1$] as $T\to 0$. If asymmetry
  is strong, $G(\epsilon_0)$ is only slightly modulated for all $T$, i.e.,
  exhibits no peak.}
\end{figure}

\subsubsection{Asymmetric barriers}

Introducing a weak asymmetry ${\rm R}_-=|{\rm R}_1-{\rm R}_2|\ll {\rm
R}\simeq {\rm R}_{1,2}$, we get for ${T}\ll \delta_-$, where
$\delta_-$ is defined in Eq.~(\ref{51}):
\be
G(\epsilon_0,{T})={{\rm R}^2 ({T}/D_r)^{2\alpha}\over
{\rm R}_-^2/4+{\rm R}^2(2\pi\epsilon_0/\Delta)^2}~.
\label{66}
\ee
The height and the width of the peak are
\be
G_p=\left({{T}\over \delta_-}\right)^{2\alpha}~,\quad
w={\Delta\over 2\pi}{{\rm R}_-\over {\rm R}}~.
\label{67}
\ee 
Thus, the asymmetry leads (Fig.~\ref{weakcond}b) to vanishing $G_p$ at
${T}\to 0$ and the width is seen to saturate with decreasing $T$, similarly to
Eq.~(\ref{63}). It is worth noting that the dependence of $G_p$ on $T$ is
non-monotonic: $G_p\propto T^{2\alpha}$ grows with increasing $T$ for $T\ll
\delta_-$, continues to grow in the range $\delta_-\ll T\ll w$ according to
$1-G_p\propto T^{-2\alpha}$, but goes down for $w\ll T\ll \Delta$, where the
correction behaves similarly to the case of symmetric barriers, $1-G_p\propto
T^{2(1-\alpha)}$. The conductance peak is narrow provided the asymmetry is
weak, ${\rm R}_-\ll {\rm R}$. If the asymmetry is strong, ${\rm R}_-\simeq
{\rm R}_1+{\rm R}_2$, the peak is completely destroyed.

\section{Conclusions}
\label{V}

In conclusion, we have thoroughly studied transport of weakly
interacting spinless electrons through a double barrier. We have
described a rich variety of different regimes depending on the
strength of the barrier, its shape, and temperature. We have developed
a fermionic RG approach to the double barrier problem, which has
enabled us to treat on an equal footing both the resonant tunneling
and resonant transmission through weak impurities. In the latter case,
we have demonstrated how the interaction-induced renormalization in
effect creates a quantum dot with tunneling barriers with a pronounced
resonance peak structure. Moreover, we have shown that even very weak
impurities, for which the renormalized transmission coefficient does
not exhibit any peak, may give a sharp peak in the conductance as a
function of gate voltage, provided that the double barrier is only
slightly asymmetric. In contrast, the resonant structure is shown to
be completely destroyed for a strongly asymmetric barrier. All the
regimes we have studied may be characterized by three different types
of behavior of the conductance peak height $G_p$ and the peak width
$w$ on temperature $T$: 

\begin{itemize}

\item[(i)] for high temperature $T\gg\Gamma({T})$,

$G_p\propto {T}^{\alpha-1}$ and $w\propto {T}$; 

\item[(ii)] for lower $T$,
depending on the shape of the barrier (whether it is symmetric or
asymmetric), either 

$G_p$ does not depend on $T$ and $w\propto
{T}^\alpha$ or 

\item[(iii)] $G_p\propto {T}^{2\alpha}$ and $w$ is constant.

\end{itemize}

One can see that none of the regimes (i--iii) supports $G_p\propto
{T}^{2\alpha-1}$ and $w\propto {T}$, as proposed in
Ref.~\onlinecite{postma01}. Further experiments would be useful to resolve the
puzzle.  Including spin and generalizing to the case of several channels
(possibly with different Fermi wavevectors) within the framework of the
present approach warrant further study.

We are grateful to V.V. Cheianov, F. Evers, L.I.~Glazman, D.L.~Maslov,
A.D.~Mirlin, Yu.V.~Nazarov, and P.~W\"olfle for interesting discussions. This
work was supported by SFB 195 and the Schwerpunktprogramm
``Quanten-Hall-Systeme" of the Deutsche Forschungsgemeinschaft, by
German-Israeli Foundation, and by Russian Foundation for Basic Research.


\begin{thebibliography}{1}
\bibitem{cutoff} In general, the ultraviolet cutoffs in the tracing
over states below and above the Fermi energy may be
different. However, possible asymmetry only leads to a renormalization
of bare couplings in the boundary conditions to
Eq.~(\ref{1}). Moreover, the paper deals with physical situations of
transport in quantum wires, in which case the cutoffs are symmetric
and $D_0$ does not exceed the Fermi energy.
\bibitem{g-ology} In terms of the classification of
Refs.~\onlinecite{solyom79,voit94}, we take into account $g_2$- and
neglect $g_4$-interactions.
\bibitem{without} The possibility of using the HF arguments in the
spirit of Refs.~\onlinecite{matveev93,yue94} is related to a peculiar
structure of a fully renormalized two-particle interaction vertex
which reduces simply to the bare interaction for a particular choice
of momenta it depends on. Within the parquet approximation, accounting
for the most divergent logarithmic terms, this property of the vertex
was realized long ago.\cite{bychkov66,dzyaloshinskii72}
\end{thebibliography}
\end{document}